\documentclass[5p]{elsarticle}
\usepackage{lineno,hyperref,graphicx,amssymb,amsmath,multirow,placeins}
\journal{Astronomy and Computing}
\biboptions{authoryear}

\begin{document}
\newcommand{\be}{\begin{equation}}
\newcommand{\ee}{\end{equation}}
\newcommand{\bq}{\begin{eqnarray}}
\newcommand{\eq}{\end{eqnarray}}
\begin{frontmatter}

\title{Abelian-Higgs cosmic string evolution with multiple GPUs}
\author[inst1,inst2,inst3]{J. R. C. C. C. Correia}\ead{Jose.Correia@astro.up.pt}
\author[inst1,inst2]{C. J. A. P. Martins\corref{cor1}}\ead{Carlos.Martins@astro.up.pt}
\address[inst1]{Centro de Astrof\'{\i}sica da Universidade do Porto, Rua das Estrelas, 4150-762 Porto, Portugal}
\address[inst2]{Instituto de Astrof\'{\i}sica e Ci\^encias do Espa\c co, CAUP, Rua das Estrelas, 4150-762 Porto, Portugal}
\address[inst3]{Faculdade de Ci\^encias, Universidade do Porto, Rua do Campo Alegre 687, 4169-007 Porto, Portugal}
\cortext[cor1]{Corresponding author}

\begin{abstract}
Topological defects form at cosmological phase transitions by the Kibble mechanism. Cosmic strings and superstrings can lead to particularly interesting astrophysical and cosmological consequences, but this study is is currently limited by the availability of accurate numerical simulations, which in turn is bottlenecked by hardware resources and computation time. Aiming to eliminate this bottleneck, in recent work we introduced and validated a GPU-accelerated evolution code for local Abelian-Higgs strings networks. While this leads to significant gains in speed, it is still limited by the physical memory available on a graphical accelerator. Here we report on a further step towards our main goal, by implementing and validating a multiple GPU extension of the earlier code, and further demonstrate its good scalability, both in terms of strong and weak scaling. A $8192^3$ production run, using $4096$ GPUs, runs in $33.2$ minutes of wall-clock time on the Piz Daint supercomputer.
\end{abstract}


\end{frontmatter}

\section{\label{intr}Introduction}
Cosmic strings and other topological defects are a generic prediction of many theories beyond the Standard Model of particle physics, being formed by means of the Kibble mechanism \citep{Kibble:1976sj}. Since the properties of these objects and their astrophysical consequences are intrinsically linked to the symmetry breaking patterns which produce them, one can think of them as fossil relics of the physical conditions in the early Universe. As such, hunting for defects in the early or recent Universe is akin to looking for evidence of specific symmetry breaking patterns. In particular a detection would indicate the presence of new physics, while a non-detection enables several constraints on theories of particle physics beyond the Standard Model. These are among the reasons why cosmic strings are a key target for next generation cosmic microwave background \citep{CORE} and gravitational wave experiments \citep{LISA}.

These searches crucially depend on the availability of high-resolution, high-dynamic range simulations of defect networks. Unfortunately, computational constraints are already a limiting factor on these searches: it is clear that the approximations introduced to compensate for the absence of data introduce systematic uncertainties comparable to statistical uncertainties \citep{Ade:2013xla,Abbott:2017mem}. Using semi-analytical models with enough degrees of freedom \citep{Martins:1996jp,Book} can mitigate this problem. However, the proper calibration of such models also requires high-resolution simulations. In order to alleviate this problem, one can attempt to exploit alternatives to the standard hardware architectures with the onus of optimization falling to the developers of the tool in question. The goal of this work is to describe how to optimize our previously developed GPU-accelerated cosmic string evolution code \citep{Correia:2018gew,Correia:2019bdl} for multiple accelerators.

Before proceeding, let us recall that there are two possible ways to simulate Abelian-Higgs cosmic string networks. In the first, one approximates the cosmic string by the action of a macroscopic (Nambu-Goto) string, which is in principle justified as long as the string effective cross-section is negligible when compared to the string length. This can be done also because the vacuum of the defect is strictly local---in the sense that fields are confined to the near-vicinity of the string core (hence confined to a world-sheet). Simulations which assume this approximation have been done by several independent groups \citep{BB,AS,FRAC,VVO,Blanco}. The advantages of these simulations are their comparatively large dynamical range and spatial resolution; the main disadvantage is that having only a one-dimensional effective action some of the key processes affecting network dynamics (such as intercommuting and loop production) are lost and have to be enforced by hand.

In the second approach to simulating strings, one places fields on a co-moving discrete lattice and evolves these fields throughout cosmic time. In the strict sense of the word one does not evolve strings, but instead evolves the fields---the strings are merely specific configurations of these fields. Examples of early Abelian-Higgs simulations in the literature include \citet{Moore:2001px} and \citet{Bevis:2006mj}. Computational limitations imply that ordinarily these simulations can only yield more modest spatial resolutions and/or dynamic ranges, but they do have one important advantage: the microscopic field dynamics is preserved, and therefore it is relatively easy to extend the simplest Abelian-Higgs case to multiple fields (including suitable couplings). This enables the numerical study of defect types that are not described by the Nambu-Goto approximation, at least without considering additional degrees of freedom, or considering instead the Kalb-Rammond action. This flexibility is reflected in the literature: one can find examples of global defect simulations like domain walls \citep{Rybak1}, monopoles \citep{Monopoles} and global strings \citep{Lopez-Eiguren:2017dmc}, as well as semilocal strings \citep{Semilocals} and even hybrid networks \citep{HindNab}.

This second approach is the one that we adopt in the present work. Until very recently both types of string simulations exploited only one architecture: Central Processing Units, in a typical distributed computing environment. Simulations which use alternative architectures (specifically, accelerator based ones) are far more scarce. So far there have been domain wall implementations by \citet{Intel} for Xeon Phi co-processors and by \citet{PhysRevE.96.043310}, and the more recent cosmic string implementation for GPUs by the present authors \citep{Correia:2018gew,Correia:2019bdl}. One limitation of our GPU implementations so far pertains to the amount of physical memory available on a graphical accelerator. A way around this involves swapping parts of the lattice from host to accelerator memory constantly, however, one expects this to negatively impact performance. In what follows we address this issue by implementing and subsequently validating an extension of our previous code for multiple accelerators, and we also quantify its scalability.

An outline of the rest of the paper is as follows. We start in Sect. \ref{lat} with a brief outline of our discretization procedure, and then proceed to describe its implementation for multiple GPUs in Sect. \ref{disc}. Our procedure for validating the code is then described in Sect. \ref{val}, following which we discuss the tests of its scalability (both in terms of strong and weak scaling) in Sect. \ref{impl}. Finally, we present some conclusions and a brief outlook in Sect.\ref{concl}.

\section{\label{lat}Discretization scheme}

Field theory simulations of defects in cosmology rely on discretizing fields on a lattice and allow their evolution to be dictated by integrators which in the continuuum limit approximate the equations of motion of the fields. Abelian-Higgs cosmic strings arise from a Lagrangian density which is invariant under local $U(1)$ transformations, which upon breaking of this symmetry eventually forms topological defects. We start by providing a brief overview of the aspects of the discretization process relevant for our code (and specifically for its multi-GPU extension), referring the reader to our previous work on the single GPU version \citep{Correia:2018gew,Correia:2019bdl} for a more detailed discussion.

The Lagrangian density has the form
\begin{equation}
	\mathcal{L}=|D_\mu \phi|^2 - \frac{\lambda}{4}(|\phi|^2 -\sigma^2)^2 - \frac{1}{4e^2}F^{\mu \nu}F_{\mu \nu}\,,
\end{equation}
where $\phi$ is a complex scalar field, the electromagnetic field tensor is given by $F_{\mu \nu} = \partial_\mu A_\nu - \partial_\nu A_\mu$, $A_\mu$ is the gauge field (where the gauge coupling $e$ has been absorbed), $D_\mu \phi$ is the gauge covariant derivative given by $D_\mu = \partial_\mu -iA_\mu$ and $\lambda$ and $e$ are coupling constants. Throughout this work we assume both the temporal gauge ($A_0 =0$) and a Friedmann-Lemaitre-Robertson-Walker flat background, described by the metric

\begin{equation}
    ds^2 = -dt^2 +a(t)^2 \bigg[ dr^2 + r^2 (d\theta^2 + \sin \theta^2 d\phi^2 ) \bigg]
\end{equation}
where $a(t)$ is a function known as scale factor and ($t, r, \theta, \phi$) are spacetime coordinates. Through a coordinate transformation this can be re-written in comoving coordinates and conformal time, which simplifies the metric to $g_{\mu \nu} = a^2 diag(-1,1,1,1)$. By standard variational principles one can obtain the equations of motion,
\begin{equation}
\ddot{\phi} + 2\frac{\dot{a}}{a}\dot{\phi} = D^jD_j\phi -\frac{a^{2}\lambda}{2} (|\phi|^2 - \sigma^2) 
\end{equation}
\begin{equation}
\dot{F}_{0j} = \partial_j F_{ij} -2a^2 e^2 Im[\phi^* D_j \phi]\,,
\end{equation} 
along with Gauss's law,
\begin{equation}
\partial_i F_{0i} = 2 a^2 e^2 Im[\phi^* \dot{\phi}]\,,
\end{equation}
where $\dot{a}$ indicates a derivative of the scale factor with respect to conformal time. 

In order to obtain the discrete form of these equations (which tell us how to update the fields at each numerical timestep) one needs to consider the principles of lattice gauge theory \citep{PhysRevD.10.2445}. Note that we will, in addition, allow both the scalar and gauge couplings to scale (this modifies the usual equations of motion and by extension also their discretizations). Both couplings are now described by,
\begin{align}
  \lambda = \lambda_0 a^{2(1-s)} 
  && 
  e = e_0 a^{(1-s)}
\end{align}
such that the value of $s$ can be used to either fix the comoving width of the defect \citep{PRS} or even to allow it to grow (for a negative $s$) and subsequently shrink as expected in the true equations of motion $s=1$ \citep{Bevis:2006mj}. In this way the defects are resolved and do not become smaller than the lattice spacing.

In terms of simulation parameters we choose lattice spacing of $\Delta x = 0.5$ and timestep size of $\Delta \eta = 0.1$, which are standard choices in the literature (\citet{Correia:2019bdl,Daverio:2015nva}, and couplings of $\lambda = 2.0$ and $e_0 = 1.0$; the latter choices ensure criticality, with a Bogomol'nyi ratio $\beta = \lambda / 2e^2 = 1.0$. These standard choices are made to enable a meaningful comparison to the most commonly simulated cases on the literature and thus allow us to validate our code, as detailed below.

The simulations start at $\eta_0=1.0$. The initial conditions consist of setting a random phase for the complex scalar field $\phi$ while other fields are set to zero. These initial conditions are an attempt to mimic the fields after a second-order phase transition (already in the broken symmetry phase). For the purpose of this paper, we note that we are not applying any period of artificial damping. However, such capability exists in the simulation and has in fact already been used to study the consequences of varying degrees of damping \citet{Correia:2020gkj}. Due to the periodic boundary conditions, the simulation ends when the horizon reaches half the box size (half-light crossing time). The final conformal time can then be computed using $0.5 \Delta x  N$ where $N$ is the box size. Altogether, these specs mean that there are 630 time-steps in a $256^3$ simulation and $1270$ time-steps in a $512^3$ simulation.

With these prescriptions one obtains a discretization based on a staggered leap-frog scheme with respect to terms of second-order in time and Crank-Nicholson with respect to terms of first order in time:
\begin{equation}
\begin{split}
(1+\delta)\Pi^{x,\eta+\frac{1}{2}} &= (1-\delta)\Pi^{x,\eta-\frac{1}{2}}+\Delta\eta  [D_j^-D_j^+\phi^{x,\eta} \\ &-\frac{\lambda_0}{2} a_\eta^{2s}(|\phi^{x,\eta}|^2-\sigma^2)\phi^{x,\eta}]
\end{split}
\end{equation}
\begin{equation}
\begin{split}
(1+\omega)E^{x,\eta+\frac{1}{2}}_i &=  (1-\omega)E^{x,\eta-\frac{1}{2}}_i +\Delta\eta [-\partial_i^- F_{ij} \\
&+ 2e_0^{2}a^{2s}_\eta Im[\phi^* D_i^+ \phi]^{x,\eta} ]
\end{split} 
\end{equation}
\begin{equation}
\phi^{x,\eta+1} = \phi^{x,\eta} + \Delta \Pi^{x,\eta+\frac{1}{2}}
\end{equation}
\begin{equation}
A^{x,\eta+1}_i = A^{x,\eta}_i + \Delta E^{x,\eta+\frac{1}{2}}_i\,,
\end{equation}
where $\omega=\delta(1-s)$, $\delta=\frac{1}{2} \alpha \frac{dlna}{dln\eta}\frac{\Delta \eta}{\eta} = \frac{1}{2} \alpha \frac{m \Delta \eta}{(1-m)\eta}$, and the last equality assumes cosmological power-law expansion rates with the scale factor
\begin{equation}\label{defm}
a\,\propto\,t^m\,\propto \eta^{m/(1-m)}\,,
\end{equation}
respectively in terms of physical and conformal time. Note that $\delta$ is responsible for the Hubble damping of the scalar field and $\omega$ is responsible for damping the gauge field. In particular, the damping of the gauge field when $s\neq 1$ is responsible for upholding the version of the previously stated Gauss's law to machine precision. We remark as well that all spatial derivatives are accurate to $\mathcal{O}(\Delta x^2)$. 

In order to completely describe the communication and memory access patterns of our code, we must also describe gauge covariant-derivatives, $D^+ \phi$, and the derivatives of the gauge field strength $\partial^- F_{ij}$ in greater detail. In order to accurately preserve gauge invariance on the lattice these two quantities can be defined using so-called link variables,
\begin{equation}
U_j^x = e^{-i A _j}\,,
\end{equation}
which describe the gauge fields on the lattice as parallel transporters of the scalar field $\phi$. With these we can properly define the modified Laplacian stencil,
\begin{equation}
D_j^-D_j^+\phi^x = \sum_j \frac{1}{\Delta x^2} [ U_j^x\phi^{x+k_j} - 2\phi_j^{x} +(U_j^{x-k_j})^* \phi^{x-k_j}]\,,
\end{equation}
and the spatial components of the gauge field strength can be defined with the real part of the following product of link variables,
\begin{align}
\Xi_{ij} &= U_i^x U_i^{x+k_j}  (U_j^{x+k_i})^* (U_j^{x})^* \\
		 &= exp[i \Delta x (\partial^+_i A'_j(x) -\partial^+_j A'_i(x) )]\,.
\end{align}

In order to validate the code and, more generally, quantitatively describe the evolution of cosmic string networks in production runs, one must compute and output at least two key network observables: the mean string separation $\xi$ and the mean velocity $v^2$. Our code can compute and output each of these variables in two separate ways. For the mean string separation these are
\begin{align}\label{diagxi}
  \xi_\mathcal{L} = \sqrt{ \frac{-\mu V}{\sum_x \mathcal{L}_x} }\,,
  && 
  \xi_W = \sqrt{\frac{\mathcal{V}}{\sum_{ij,x} W_{ij,x}}}\,.
\end{align}
The first one comes from \citet{Bevis:2006mj} and is based on the Lagrangian being strongly negatively peaked at the string core while approaching zero away from the string. The second one comes from the lattice based winding from \citet{Kajantie:1998bg}.
We also use two different estimators to compute the mean squared velocity,
\begin{align}\label{diagvel}
  \langle v^2\rangle_{\phi} = \frac{2R}{1+R}\,,
  &&
  \langle v^2\rangle_{\omega} = \frac{1}{2} \bigg( 1+3\frac{\sum_x p_x \mathcal{W}_x}{\sum_x \rho_x  \mathcal{W}_x} \bigg)\,,
\end{align}
where $R$ is the ratio,
\begin{equation}
R = \frac{\sum_x |\Pi|^2 \mathcal{L}}{\sum_{x,i} |D^+_{x,i} \phi|^2 \mathcal{L} }
\end{equation}
The first estimator comes from \citet{Hindmarsh:2008dw} and is based on considering a boosted static string. A complete derivation can be found in \citet{Hindmarsh:2017qff}. The second estimator is based on the equation of state parameter and has also been used in \citet{Hindmarsh:2017qff}. Note that these are not the only velocity estimators, but according to \citet{Hindmarsh:2017qff} these are the two estimators which are in better agreement with the expected velocity of an oscilating string in Minkowski space.

Deep in eras where the scale factor evolves according to Eq. \ref{defm} (that is, with a constant expansion rate $m$), we expect these observables to exhibit scale-invariant behaviour \citep{Book}: the string separation should grow linearly with time, $\xi\propto\eta$, and the mean velocity should be constant, $v\propto const$. Note that the proportionality factors for different expansion rates $m$ depend on $m$ itself and also on other parameters, in a way that is quantitatively described by the analytic velocity-dependent one-scale model \citep{Martins:1996jp,Book}. A recent accurate calibration of this model has been presented in \citet{Correia:2019bdl}.

\section{\label{disc}Extension to multiple accelerators}

In order to enable taking a large lattice and dividing it among multiple accelerators, such that they can all partake in evolving the fields, we must consider that due to the modified Laplacian stencil and the derivative of the link variables, such field values ($\phi, A$) must be communicated between sub-domains. In order to allow such extension to multiple nodes in a network (as is standard in most high performance computing facilities) we use the Message-Passing-Interface (MPI) to facilitate communications. Throughout this work we assume a 3D decomposition.

\begin{figure*}
\centering
\includegraphics[width=\textwidth]{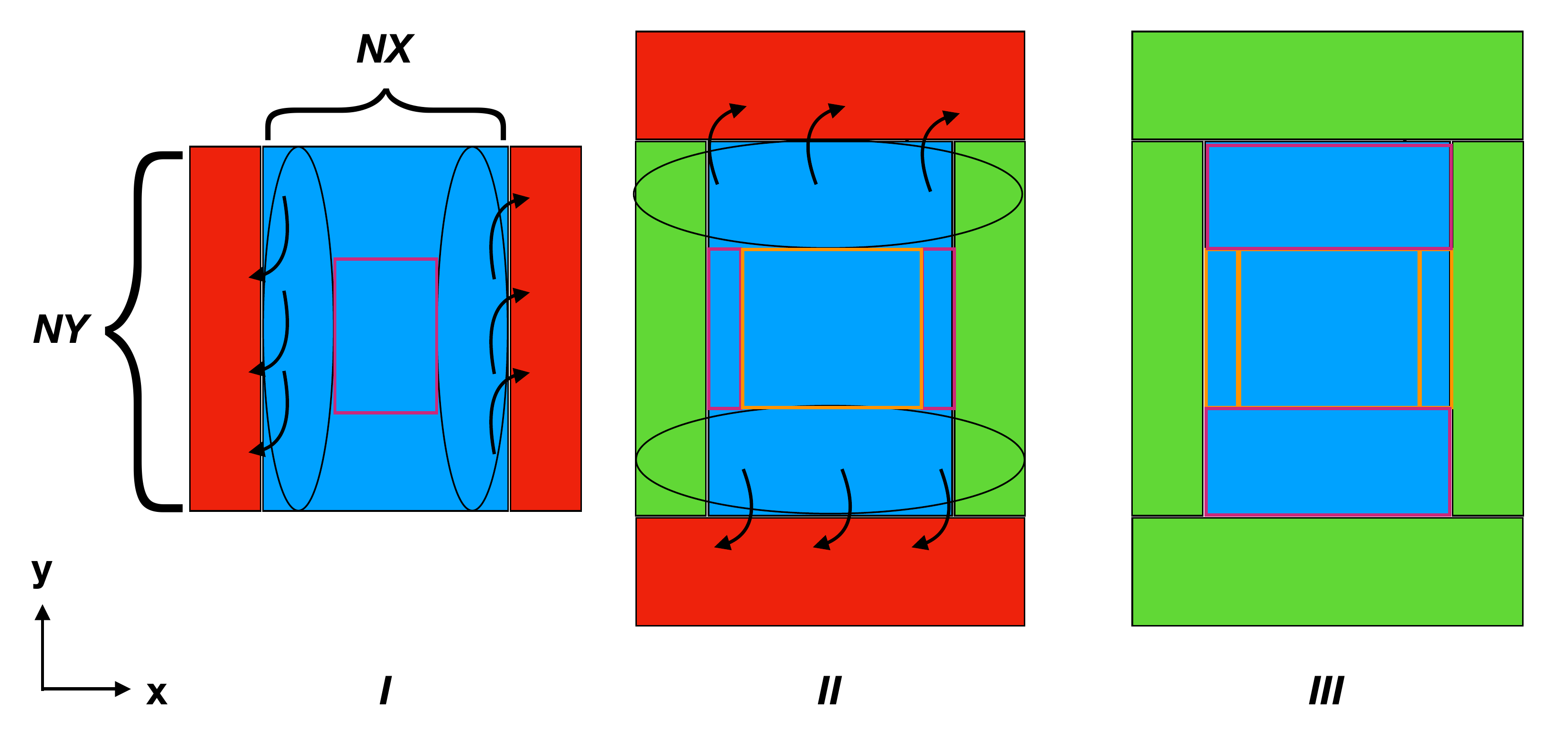}
\caption{The packing procedure along two different directions. Blue represents the core of each domain (of size $NX \times NY$), red represents the buffers being filled with appropriate values to send to neighboring sub-domains and green represents an already received buffer. In the left panel, the buffer values come only from the blue inner core. After communication has taken place in this first direction, we can unpack the received buffer into the boundaries of the sub-domain. Once done, we can start packing the communication buffers for the next direction. This involves using not only the blue inner core but the freshly unpacked boundaries (in green). The pink boxes indicate domain areas where we update fields $E$ and $\phi$ either as the packing procedure begins (left and middle panels) or after all communication has taken place (right panel) whereas orange indicates these areas have already been updated.}
\label{fig1}
\end{figure*}

In the Compute Unified Device Architecture (CUDA), all code to be executed by a graphical accelerator is contained in functions denoted as kernels. In order to implement the 3D decomposition, in addition to the kernels which evolve field quantities, we add kernels that pack outer values of the core of each sub-domain into additional buffers which are then sent to neighboring sub-domains via \textit{Isend} (from MPI). After both \textit{Isend/Irecv} complete (a \textit{WaitAll} barrier is necessary to ensure that all communication is complete) we use similar unpacking kernels to place the contents of received buffers into the boundaries of each sub-domain. Note that launching CUDA kernels from the host is non-blocking (relative to the host) which is why we must include a \textit{CUDAStreamSynchronize}, which ensures the packing kernel and all kernels before it have completed in time, in appropriate places. Note that a CUDA Stream is a sequence of commands launched in order (in this case a sequence of kernels). In order to obtain maximal bandwidth and minimal latency the communicated buffers are allocated in Unified Memory and Remote direct memory access is allowed, as noted in standard good practices \citep{pracebest}.

It must be emphasized that some key requirements must be fulfilled in order to correctly satisfy all boundary conditions. For example, due to the diagonal terms of the gauge field $A_i$ (needed to compute the derivative of the gauge field strength) we must also pay attention to "corners". The way to correctly handle this is to use the so-called "diagonal trick" which means the corners along a given direction come from values exchanged from the previous communication in another direction. This dependency of exchanges in one direction upon preceding exchanges implies that communication must proceed in a given specific order. In our case this means communication must proceed as follows:

\begin{enumerate}
\item Pack the values to be sent to neighbors along X (outer part of the inner $NX \times NY \times NZ$ part of the domain);
\item Send packed buffers to neighboring sub-domains;
\item Unpack received values into boundaries in the X direction;
\item Pack the values from the outer cells of the inner $NX \times NY \times NZ$ along with values received from the previous exchange (to ensure corners are appropriately handled), to be sent to neighbors in the Y direction;
\item Exchange packed buffers in the Y-direction;
\item Unpack received buffers into the boundaries of the sub-domain;
\item Pack the values from not only the inner core from the sub-domain but also from the two previous exchanges;
\item Exchange packed buffers in the Z-direction;
\item Unpack received buffers into boundaries.
\end{enumerate}

Having done this we can choose to perform the update of $E$ and $\Pi$. However, in order to obtain Weak scaling above $90\%$ for thousands of GPUs we must also consider overlapping the compute and communication work. In order to compute overlap we can update the inner core of each sub-domain while we start packing the values for communication along X. Note that the outermost cells of this inner core require values from the boundaries which are still being communicated. This means that the size of the inner core we update must be $(NX-2) \times (NY-2) \times (NZ-2)$ (while for communication it remains $NX \times NY \times NZ$. After exchange in the X-direction is completed, we can begin updating the outer part of the sub-domain along X (given that the necessary boundary is not available), while communication is completed in the Y-direction. This proceeds until all boundaries are exchanged and $E$ and $\Pi$ are updated everywhere. At such a point, we can simply update $\phi$ and $A$.

A schematic view of this is presented, for the simpler case of 2 dimensions, in Figure \ref{fig1}. Note that here we can also make use of multiple CUDA streams (asynchronous with respect to each other) in order to allow overlap between the compute kernels and the pack/unpack kernels (one stream per each pack/unpack kernel). This also means that the correct dependencies between streams must be enforced. This can be done using a combination of \textit{cudaEventRecord} (which signals the completion of a kernel in a given stream) and \textit{cudaStreamWaitEvent} (which can force a stream to wait for completion of a given event in another stream).

\begin{table*}
  \begin{center}
    \begin{tabular}{|c|c|c|c|c|c|c|c|c|}
    \hline
      \textbf{Size} & \textbf{m} & \textbf{$\dot{\xi_\mathcal{L}}$} & \textbf{$\dot{\xi_W}$} & \textbf{$\langle v^2 \rangle_\omega$} & \textbf{$\langle v^2 \rangle_\phi$} & Reference\\
      \hline
$1024^3$ & 1/2 & $0.280\pm0.023$ & $0.282\pm0.026$ & $0.306\pm0.003$ & $0.272\pm0.002$ & This work\\
$2048^3$ & 1/2 & $0.268\pm0.011$ & $0.267\pm0.010$ & $0.312\pm0.001$ & $0.283\pm0.001$ & This work\\
$4096^3$ & 1/2 & $0.253\pm0.007$ & $0.251\pm0.006$ & $0.308\pm0.002$ & $0.282\pm0.001$ & This work\\
      \hline
$512^3$  & 1/2 & $0.30\pm0.02$ & $0.32\pm0.03$ & $0.32\pm0.01$ & $0.31\pm0.01$ & \citet{Correia:2018gew} \\
$512^3$  & 1/2 & $0.31\pm0.02$ & - & - & - & \citet{Bevis:2006mj} \\
$1024^3$  & 1/2 & - & $0.26\pm0.02$ & - & - & \citet{Bevis:2010gj} \\
$4096^3$  & 1/2 & $0.234\pm0.006$ & $0.244\pm0.005$ & - & - & \citet{Daverio:2015nva} \\
      \hline
$1024^3$ & 2/3 & $0.279\pm0.016$ & $0.285\pm0.017$ & $0.255\pm0.003$ & $0.228\pm0.004$ & This work\\
$2048^3$ & 2/3 & $0.256\pm0.006$ & $0.257\pm0.005$ & $0.264\pm0.001$ & $0.240\pm0.001$ & This work\\
$4096^3$ & 2/3 & $0.252\pm0.010$ & $0.250\pm0.009$ & $0.265\pm0.001$ & $0.243\pm0.001$ & This work\\
      \hline
$512^3$  & 2/3 & $0.29\pm0.01$ & $0.29\pm0.02$ & $0.27\pm0.01$ & $0.25\pm0.01$ & \citet{Correia:2018gew} \\
$512^3$  & 2/3 & $0.30\pm0.01$ & - & - & - & \citet{Bevis:2006mj} \\
$1024^3$  & 2/3 & - & $0.28\pm0.01$ & - & - & \citet{Bevis:2010gj} \\
$4096^3$  & 2/3 & $0.235\pm0.008$ & $0.247\pm0.008$ & - & - & \citet{Daverio:2015nva} \\
		\hline
  \end{tabular}
\caption{The asymptotic rate of change of the mean string separation $\xi$ and the mean velocity squared $\langle v^2\rangle$ for the estimators defined in the text, in the radiation and matter eras ($m=1$ and $m=2$ respectively), for our simulations with box sizes of $4096^3$, $2048^3$ and $1024^3$, using $4096$, $512$ and $64$ GPUs respectively. The error bars are the statistical uncertainties from averages of 20 runs with different initial conditions. For comparison we show the results reported in \protect\citet{Correia:2018gew} from the single GPU code (for averages of twelve $512^3$ simulations) as well as results from simulations with CPU-based codes. The range of timesteps used for each fit to the GPU simulations is respectively $[517,1023.5]$, $[300.5,511.5]$, $[100.5,255.5]$, $[80,128]$ for the $4096^3$, $2048^3$, $1024^3$ and $512^3$ simulations.\label{tab1}}
    \end{center}
\end{table*}

\section{\label{val}Code validation}

In order to verify that the simulation boxes evolved by the new code behave as expected, one must compare the numerically measured physical estimators (the slope of the time dependence of the mean string separation and the asymptotic mean velocity squared) to the values available in the literature, including those previously obtained (at different box sizes) with the single GPU version of the code, which has been previously validated \citep{Correia:2018gew}. We will do this with constant co-moving width simulations in the canonical radiation and matter dominated epochs, since these are the most common in the literature. 

\begin{figure*}
\centering
  \includegraphics[width=\columnwidth]{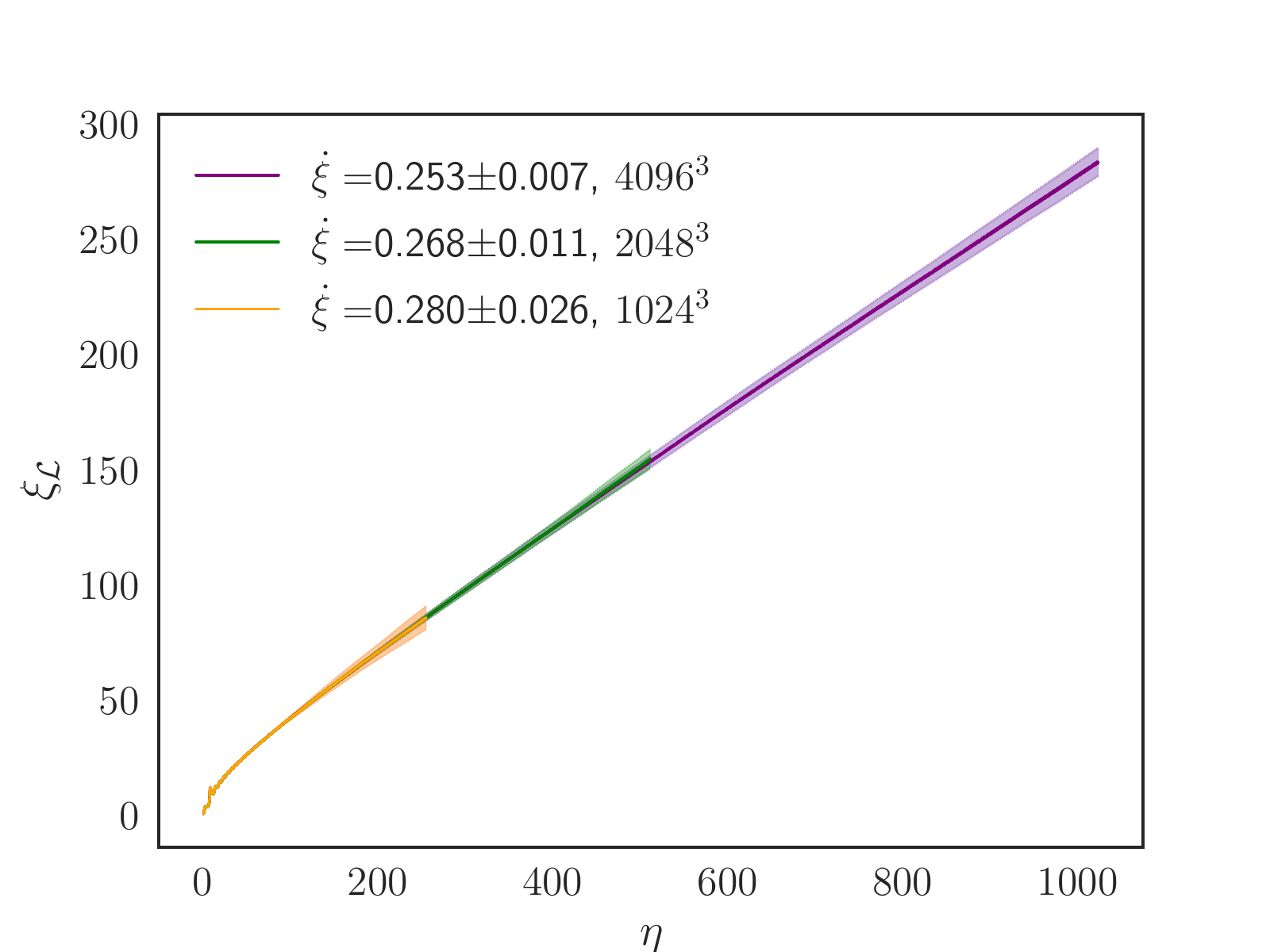}
  \includegraphics[width=\columnwidth]{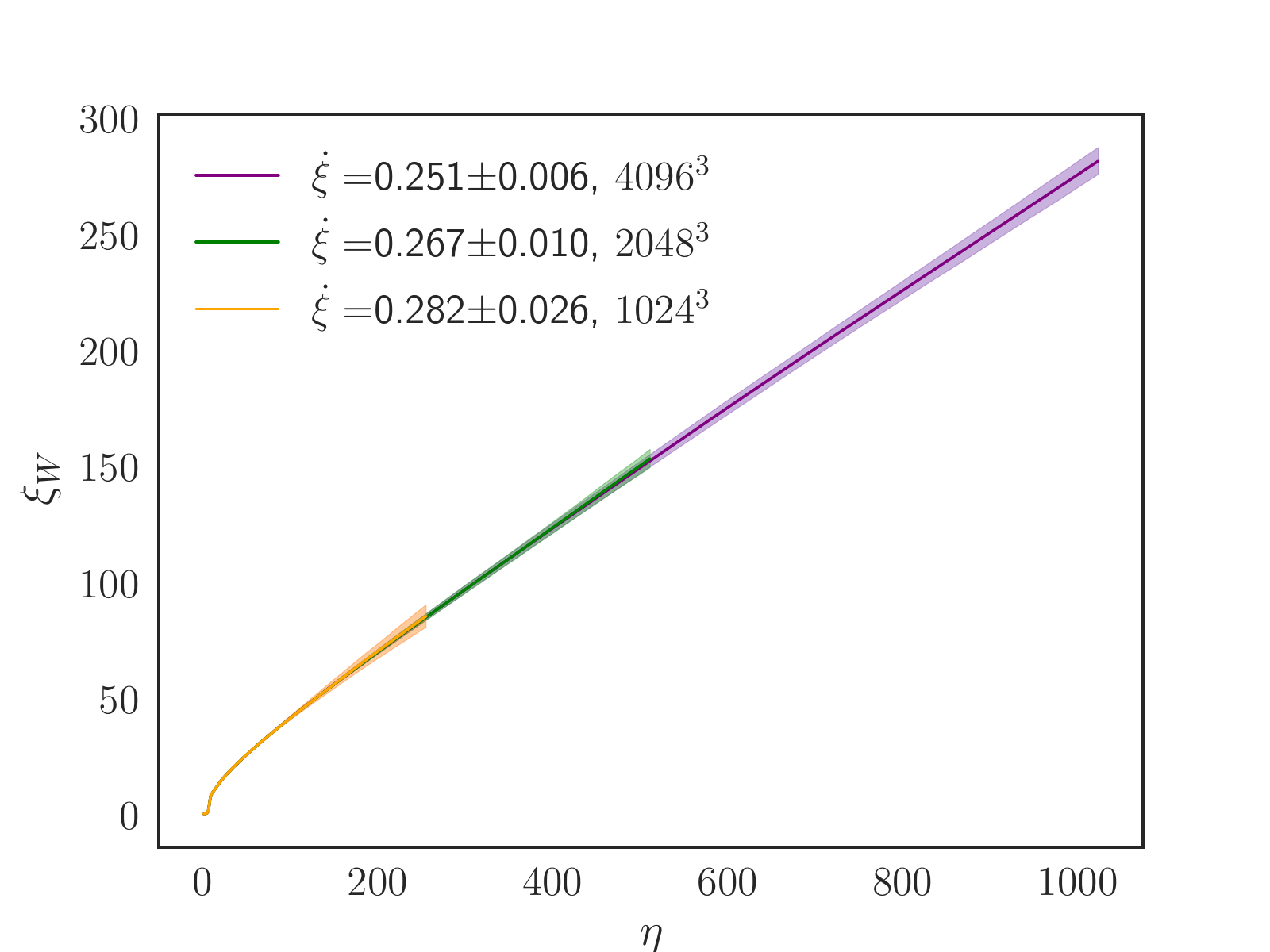}
  \includegraphics[width=\columnwidth]{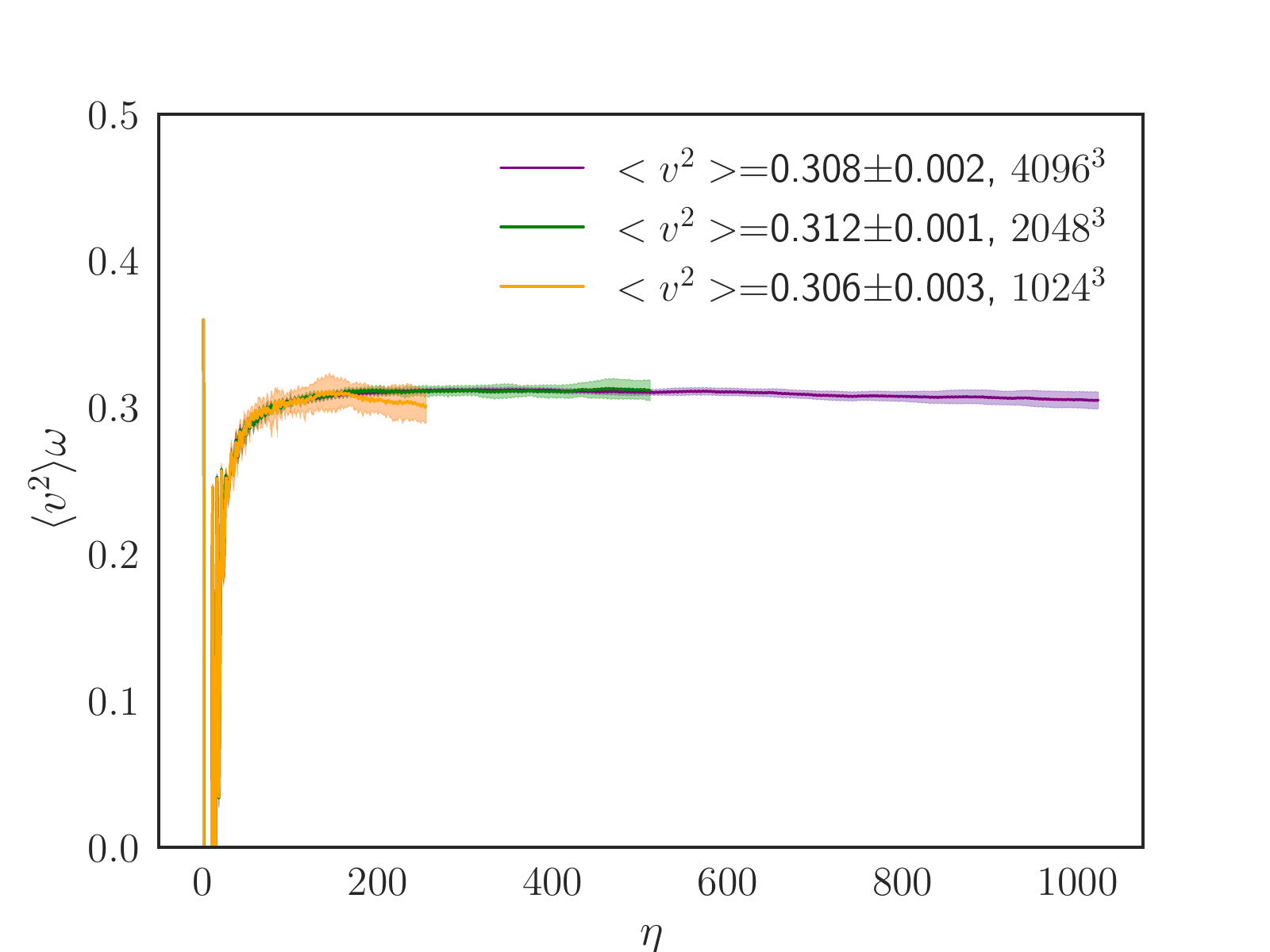}
  \includegraphics[width=\columnwidth]{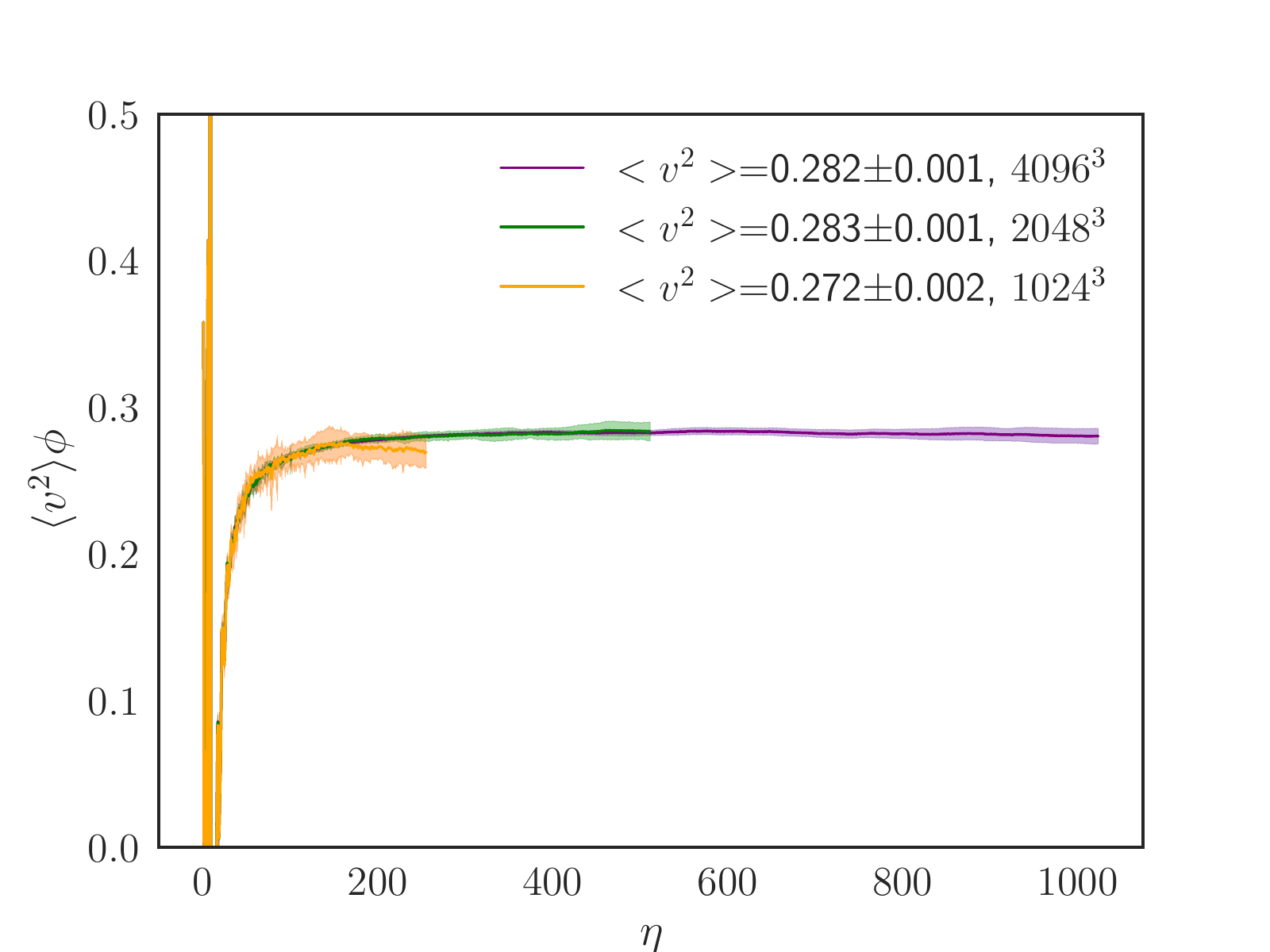}
  \caption{The evolution of the four relevant average network estimators, defined in Eqs. \ref{diagxi} and \ref{diagvel}, for the average of 20 runs in the radiation-dominated epoch ($m=1/2$), with lattice sizes of $4096^3$, $2048^3$ and $1024^3$, using $4096$, $512$ and $64$ GPUs respectively. We assume constant co-moving width throughout.}
  \label{fig2}
\end{figure*}

\begin{figure*}
\centering
  \includegraphics[width=\columnwidth]{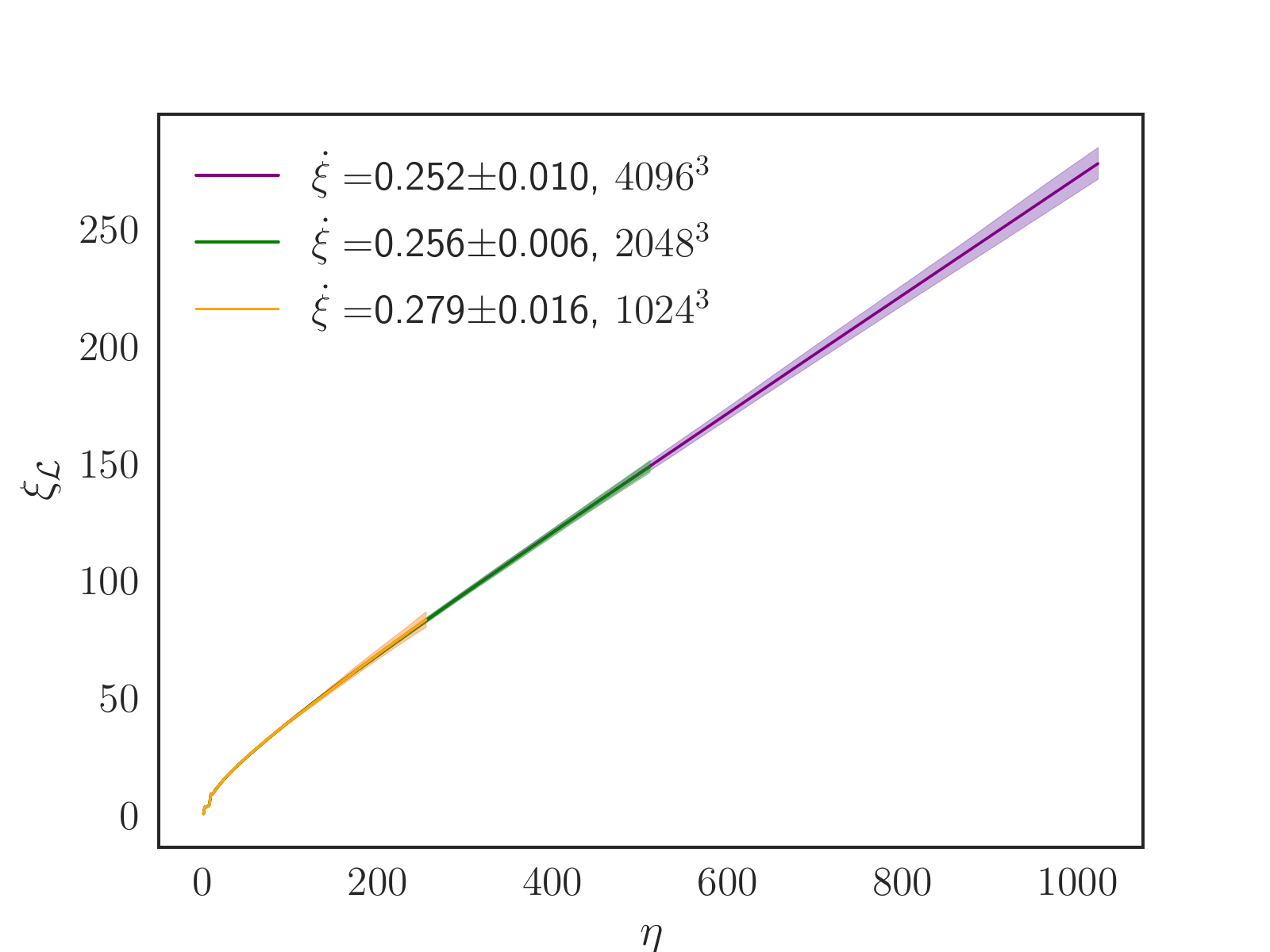}
  \includegraphics[width=\columnwidth]{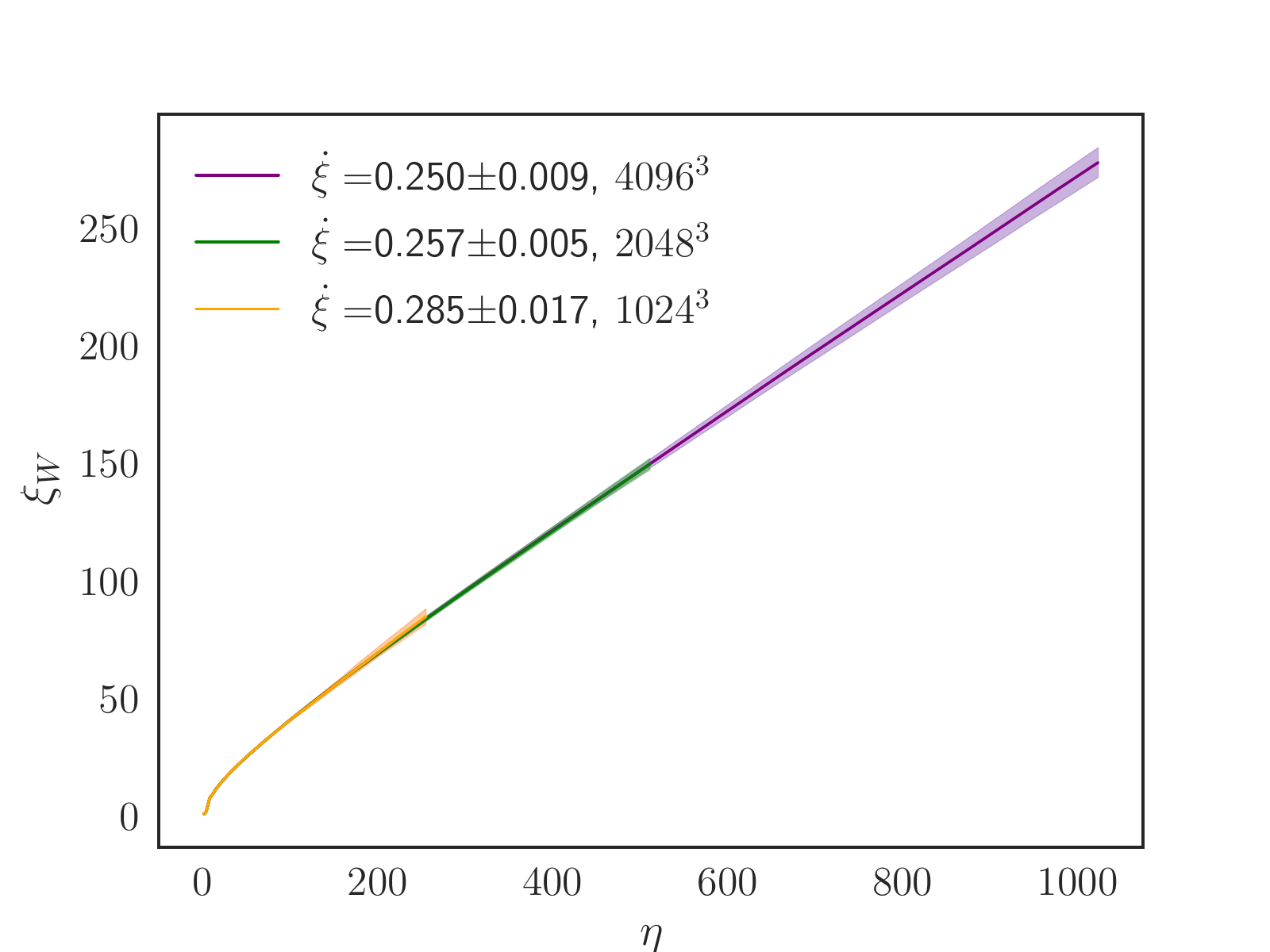}
  \includegraphics[width=\columnwidth]{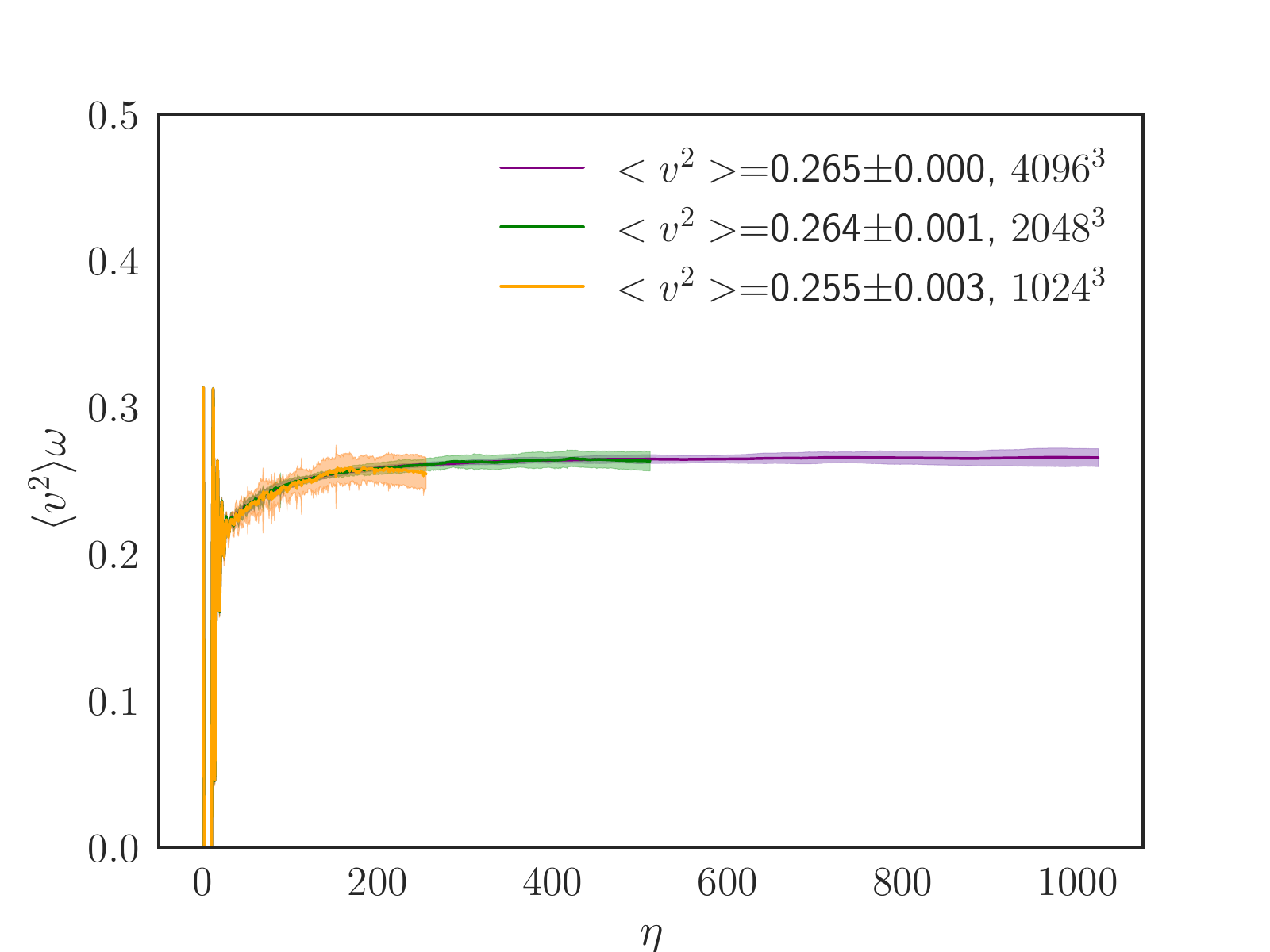}
  \includegraphics[width=\columnwidth]{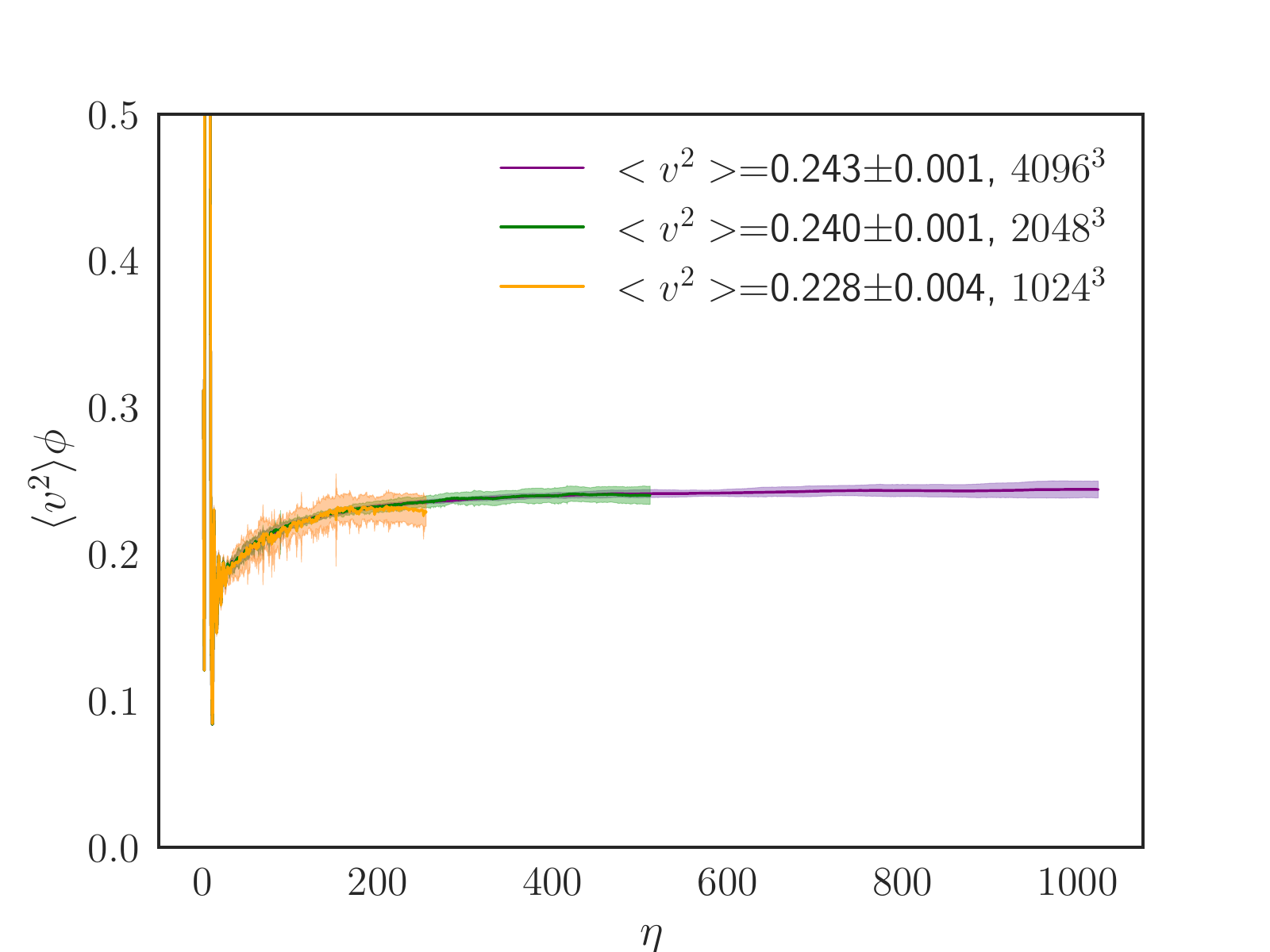}
  \caption{Same as Fig. \protect\ref{fig2}, for the matter-dominated epoch ($m=2/3$).}
  \label{fig3}
\end{figure*}

In order to conduct this test, we simulate $20$ runs for each expansion rate, using the same initial condition for both matter and radiation. Using roughly the final third of the time-steps we obtain the slopes of each curve and the average and standard deviation are then used to extract the asymptotic quantities and their one sigma statistical uncertainties. The results of this validation test are summarized in Table \ref{tab1} and also in Figures \ref{fig2} and \ref{fig3} (respectively for the radiation and matter dominated eras), and in a nutshell the new code is in agreement with the results in the literature, given the reported  uncertainties. 

Regarding the mean string separation, our previously obtained values at $512^3$ were in agreement with the values obtained by \citet{Bevis:2006mj} with the same box size. The larger simulations in the current work confirm the slight drift of the scaling value of $\dot{\xi}$ to lower values, as can be seen in \citet{Bevis:2010gj} for $1024^3$ boxes and \citet{Daverio:2015nva} at $4096^3$ (see also \citet{Hindmarsh:2017qff}). This slow drift may be due to the fact that a higher (spatial) resolution affects the main energy loss mechanims for the network (loop production and scalar and gauge radiation) in slightly different ways, which in turn impacts the string network density. A detailed exploration of this hypothesis, in the context of the recently improved and calibrated velocity-dependent one scale model \citep{Correia:2019bdl} is in progress. We also note that the two independent estimators for the mean string separation, defined in Eq. \ref{diagxi}, lead to fully consistent values for $\dot{\xi_\mathcal{L}}$ and $\dot{\xi_W}$.

As for the average velocity squared, our previous work \citep{Correia:2018gew,Correia:2019bdl} using the estimators of \citet{Hindmarsh:2017qff} had already established qualitative agreement with the values in the literature, up to and including $4096^3$ simulations. Here this agreement continues. Note that in the case of the velocities there is no statistically significant drift in the scaling value as a function of the box size. On the other hand, and in agreement both with \citet{Hindmarsh:2017qff} and with our earlier $512^3$ study, our present analysis confirms that the velocity estimator based on the gradient on $\phi$ leads to values that are consistently lower than those of the equation of state estimator, by about ten per cent at all box sizes.

\section{\label{impl}Performance}

All the performance tests we report herewith were performed at Piz Daint, the largest supercomputer in Europe. At the time of the measurements, this facility contains $5704$ nodes each equipped with with one NVIDIA Tesla P100. All benchmarks are performed assuming the evolution of a local string network as described in the previous sections.

To closely mimic a typical use case (in other words, a typical production run), we choose to compute the Lagrangian based mean string separation and the mean velocity squared estimated from $\phi$ and its conjugate field $\Pi$, weighted with the Lagrangian. The computation of these network averaged quantities occurs at every $5$ time-steps. The initial conditions are generated at random in each case.

We will test our code performance in terms of two metrics: strong and weak scaling. To define each, let's consider two possible test cases: the first is an application which is compute bound and requires an enormous amount of wall-clock time and the second is an application which is limited by the amount of available memory. In the first case, given a constant problem size, one wishes to increase the number of processors used for the computation, thus decreasing the workload for each processor and overall wall-clock time. Characterizing this behavior will tell us how much we can afford to sub-divide our simulation grid for a specific problem size in terms of performance gains. This scalability diagnostic is known as strong scaling. For the second application, we wish to target larger and larger problem sizes. So for this type of scalability, we keep the workload per process constant and both the problem size and number of computational elements increase. This diagnostic is known as weak scaling.

Both strong and weak scaling are arguably relevant metrics for the problem at hand, but in pragmatic terms the weak scaling is the most relevant one. The limiting factor in contemporary simulations is clearly the amount of available memory, and indeed one must often extrapolate from these relatively small simulations to cosmologically relevant scales. Targeting larger and larger simulation sizes (and therefore larger dynamic ranges) would lessen this problem. In particular, doubling the box size along each dimension, besides the obvious increases in volume (by a factor of 8) and in dynamic range (by a factor of 2), also increases the range of physical scales between string thickness and horizon size that can be probed. In our case the strong scaling would only become critical if the total wall-clock times of the simulation were much larger.

For simulations of cosmic string networks (or indeed those of other cosmological defects) the dynamic range of the simulation increases with the size of the whole box, being proportional to the length of the smallest box side $N$ (for any cubic simulation box $N^3$). Given this feature of string simulations, for the weak scaling diagnostic we quantify the time taken to evolve the lattice a number of time-steps equal to the number of time-steps required to evolve the smallest simulation box we considered, which has a size of $256^3$. With our choices for other numerical parameters, described above, this corresponds to a total of $630$ time steps.

\begin{figure*}
\centering
  \includegraphics[width=\columnwidth]{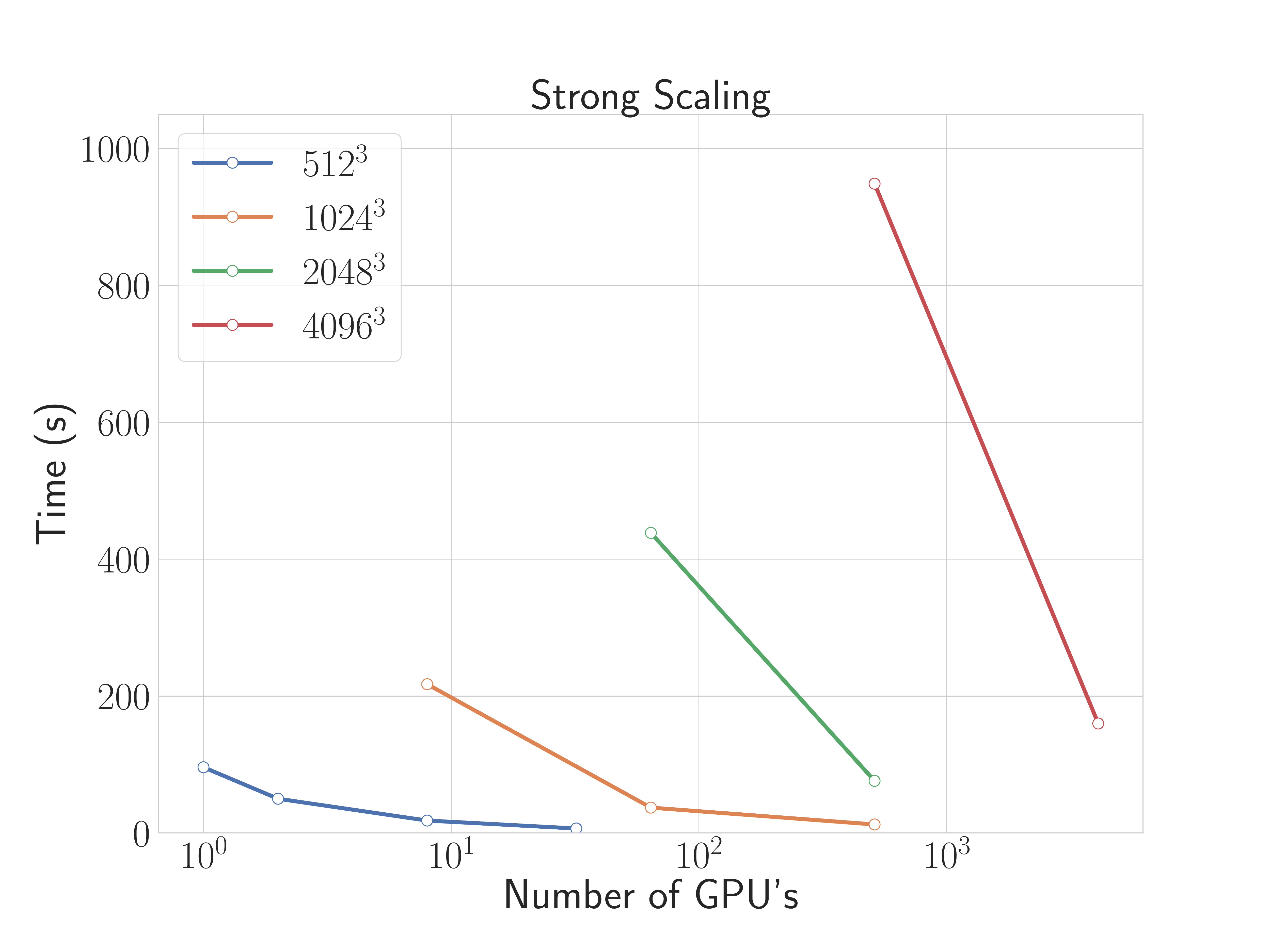}
  \includegraphics[width=\columnwidth]{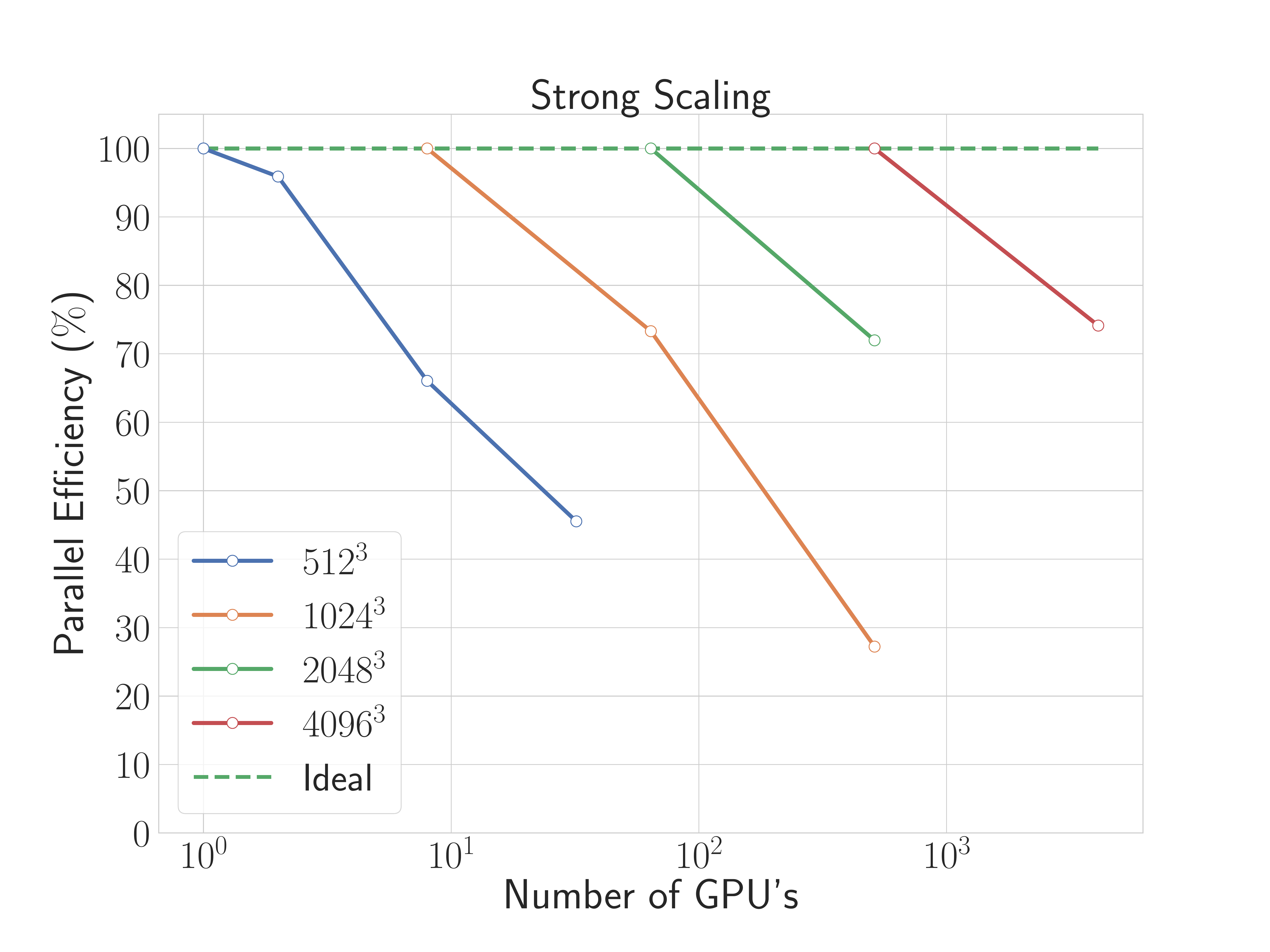}
  \includegraphics[width=\columnwidth]{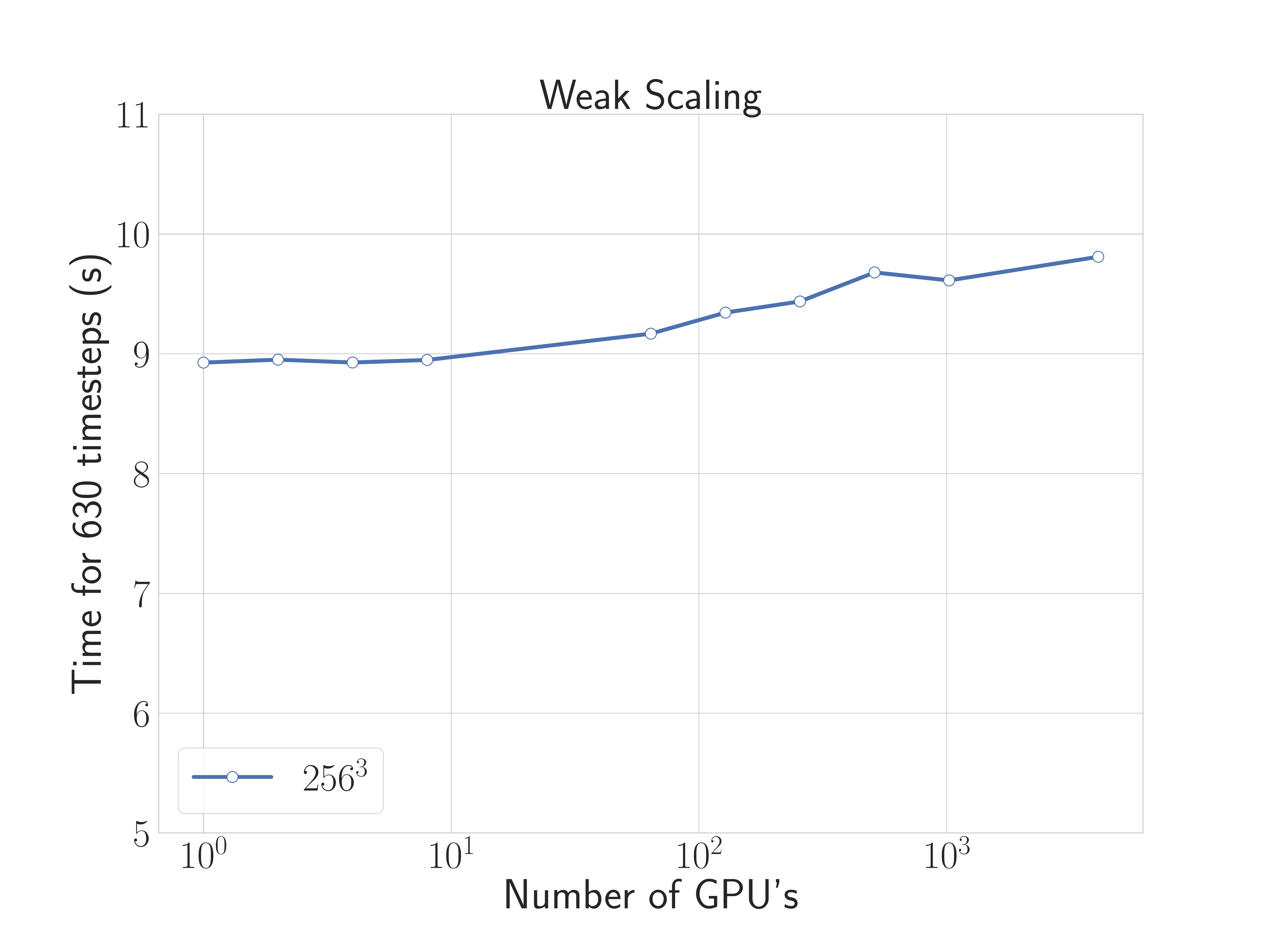}
  \includegraphics[width=\columnwidth]{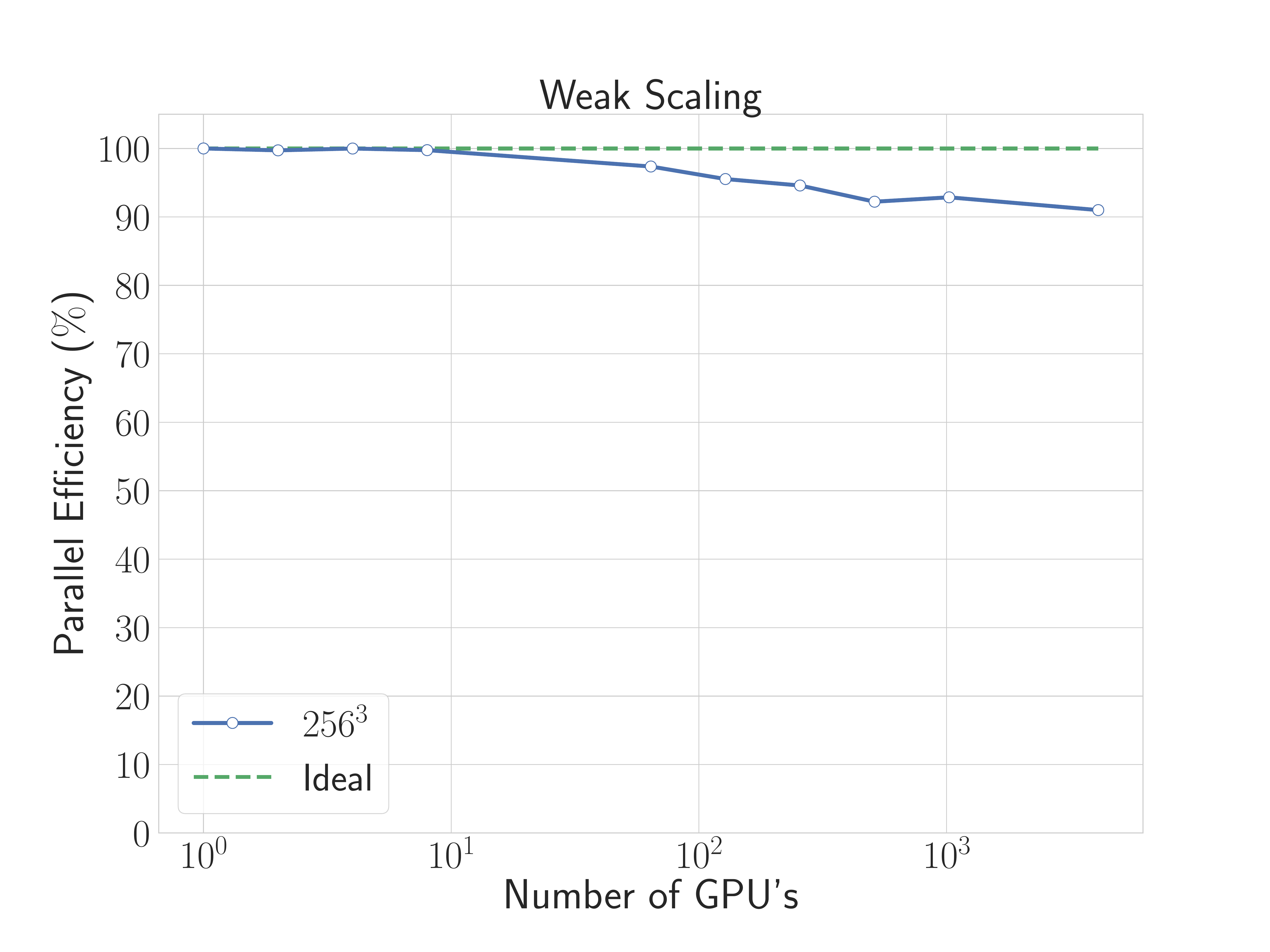}
  \includegraphics[width=\columnwidth]{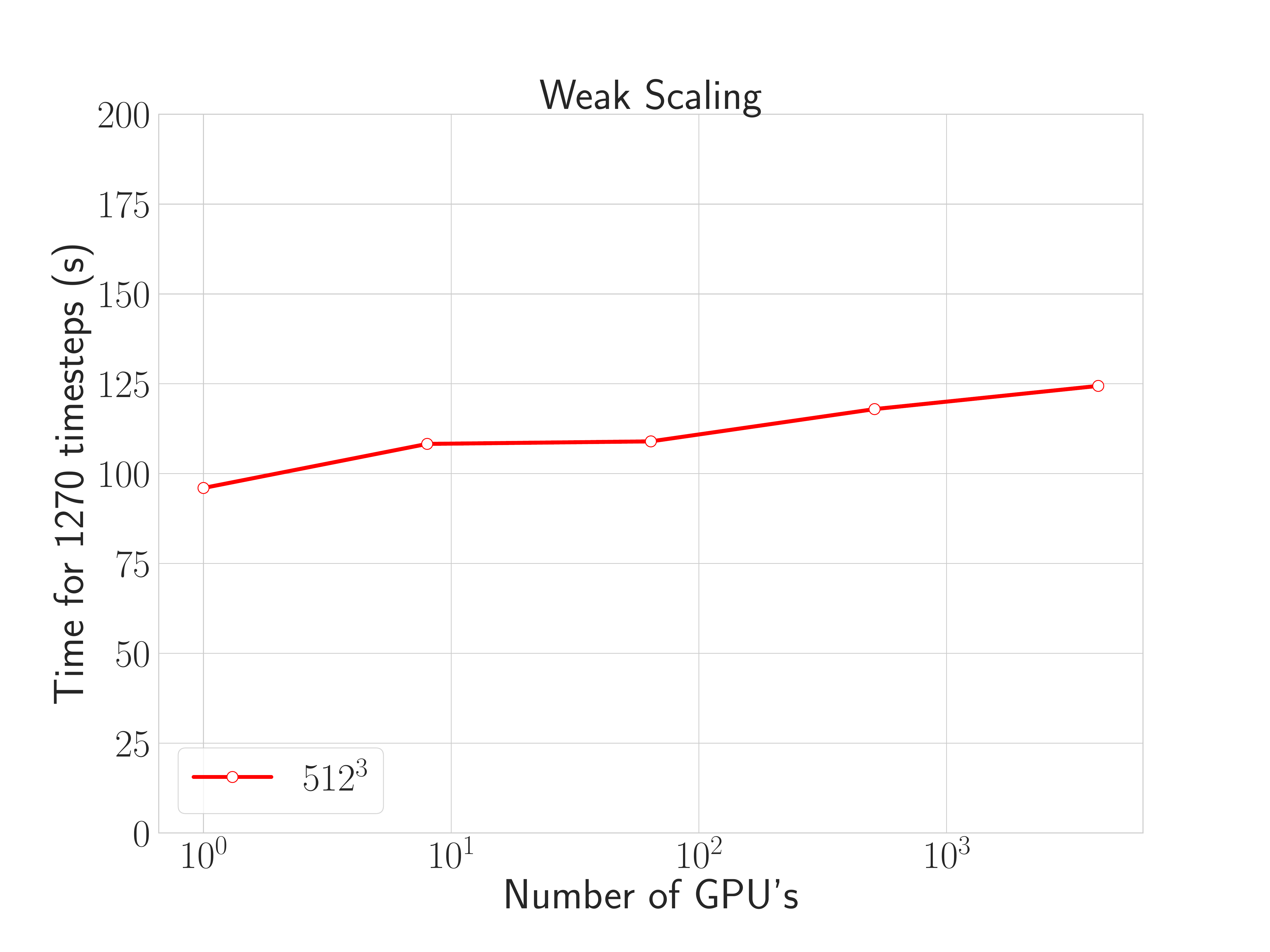}
  \includegraphics[width=\columnwidth]{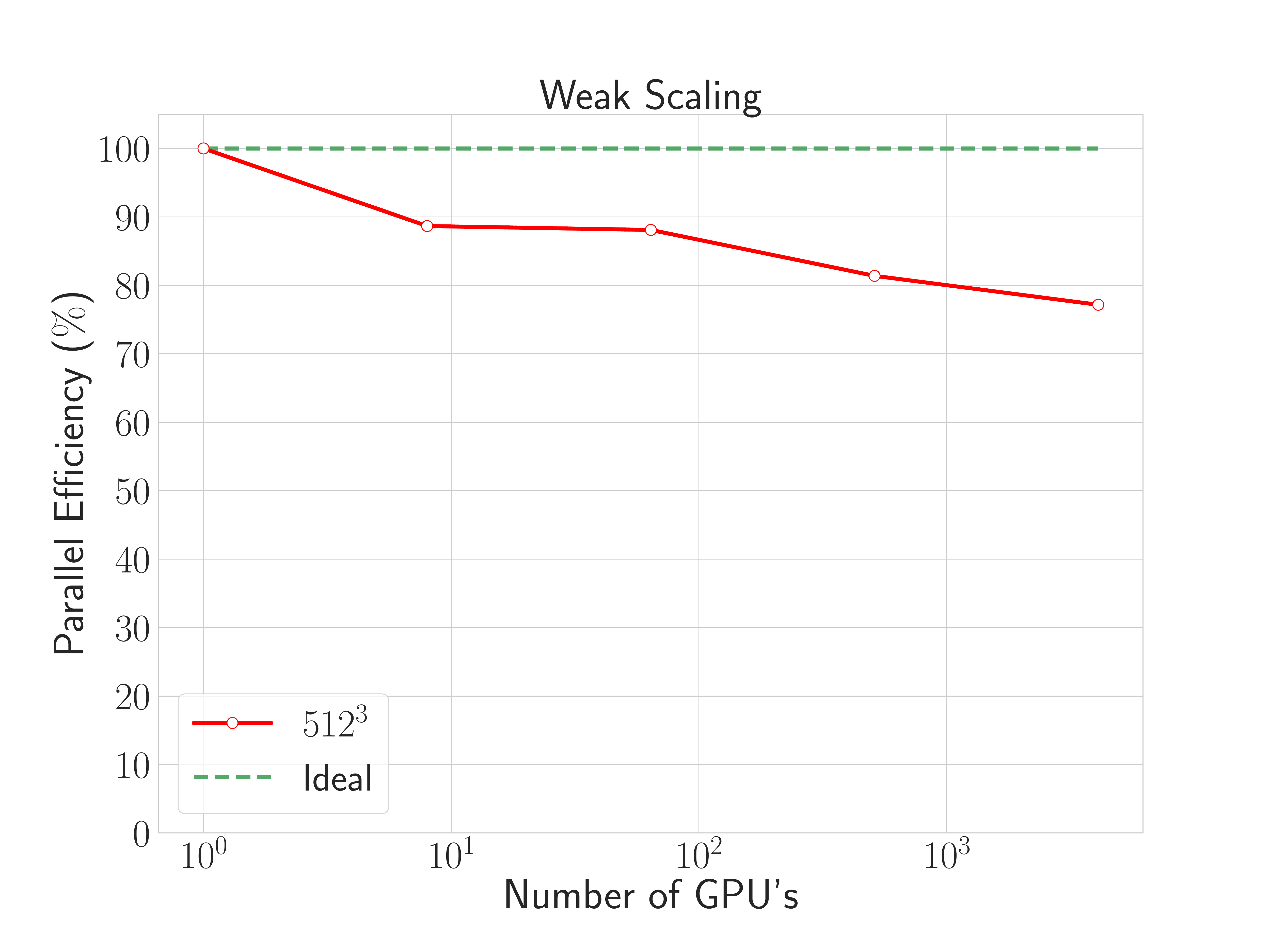}
  \caption{Performance indicators for our multiple GPU code; strong scaling is shown in the top set of panels, while weak scaling can be seen in the middle and bottom ones. The left-hand side panels show wall-clock time for a full-run (for the strong scaling plot) or the amount of wall-clock time necessary to complete $630$ time-steps (middle panels for the weak scaling plot at $256^3$) or to complete $1270$ timesteps (bottom panels at $512^3$). The corresponding parallel efficiencies relative to reference cases as defined in the text (see e.g. Eq. \protect\ref{eq:eff} for strong scaling) are presented on the right hand side panels.}
  \label{fig4}
\end{figure*}

We characterize both the weak and strong scaling using a speed-up factor, $S$, and a parallel efficiency, $E$. Both are calculated in comparison to a reference wall-clock time, denoted $t_{ref}$. In strong scaling this corresponds to the wall-clock time necessary to fully evolve the full dynamic range with the smallest number of GPUs (where a box of size $N^3$ can be fitted). The speed-up is given by
\begin{equation}
	S = \frac{t_{ref}}{t_{n}}\,,
\end{equation}
where $t_n$ is the wall-clock time taken when running with $n$ GPUs. For weak-scaling we can also use this formula, with $t_{ref}$ being the time taken for a $256^3$ simulation with one GPU and $t_{n}$ the time taken to perform an $n$-times larger simulation (with the same number of timesteps) in $n$ GPU's. Weak scaling parallel efficiency is then the speed-up as a percentage. For strong scaling the parallel efficiency is re-scaled with the number of GPUs which the reference run uses, $n_{ref}$, that is
\begin{equation}
  E_{strong} = \frac{n_{ref} t_{ref}}{n t_{n}}\,.
  \label{eq:eff}
\end{equation}
where, as above, $t_n$ is the time taken to run the simulation with $n$ GPU's.
With these defined we are  in a position to describe the scalability of our application. For strong scaling there is an obvious point beyond which no useful scaling can be obtained. While we are unaware of any consensus on the definition of useful scaling, in this manuscript we assume useful scaling to only exist above $50\%$ efficiency. Specifically, in our case this point ensues when the sub-domain size becomes too small. This is evidenced by the low parallel efficiencies seen in Table \ref{tab2} and in the top panels of Figure \ref{fig4} when approaching a sub-domain size of $128^3$. This is something relatively common in most multi-GPU implementations, at least from a cursory overview of the literature \citep{gmd-11-1665-2018,Potter:2016ttn}.

This behaviour stems from two reasons. The first reason is the amount of communications relative to the execution of CUDA kernels: not even the overlap is sufficient for cleverly hiding this cost for extremely small sub-domains. The second reason which contributes to this behaviour is the amount of latency from launching CUDA kernels.

\begin{table*}
  \begin{center}
    \begin{tabular}{c|c|c|c|c|c}
      \textbf{Box size} & \textbf{Number of GPU's} & \textbf{Domain decomposition} & \textbf{Wall-clock time} & \textbf{Speed-Up} & \textbf{Efficiency}\\
      & & (x,y,z) & (s) &  &  (\%) \\
      \hline
      $512^3$& 1  & (1,1,1) & 96.0 & -    & - \\
      		 & 2  & (1,1,2) & 50.1 & 1.92 & 95.9  \\
      		 & 8  & (2,2,2) & 18.2 & 5.16 & 66.0  \\
      		 & 32 & (2,4,4) & 6.59 & 14.57 & 45.5  \\
      \hline
     $1024^3$& 8   & (2,2,2) & 217.39 & -    & -    \\
      		 & 64  & (4,4,4) & 37.06 & 5.87  & 73.3     \\
      		 & 512 & (8,8,8) & 12.48 & 17.41 & 27.2     \\
      \hline
     $2048^3$& 64   & (4,4,4) & 438.45 & -    & - \\
      		 & 512  & (8,8,8) & 76.15 &  5.76 & 72.0 \\
      \hline
     $4096^3$& 512   & (8,8,8) & 948.52 & -    & - \\
             & 4096   & (16,16,16) & 156.96 &  6.04 & 74.3 \\
      \hline
     $8192^3$& 4096   & (16,16,16) & 1990.51 & -    & - \\
  \end{tabular}
    \caption{Strong scaling measurements for different lattice sizes reported in wall clock time to fully simulate a network from start to finish. We also present the speed-up and a parallel efficiency, both relative to the reference measurement, which uses the smallest number of GPU's where the simulation will fit. \label{tab2}}
    \end{center}
\end{table*}

\begin{table*}
  \begin{center}
    \begin{tabular}{c|c|c|c|c|c} 
      \textbf{Box size} & \textbf{Number of GPU's} & \textbf{Domain decomposition} & \textbf{Wall-clock time} & \textbf{Speed-Up} & \textbf{Efficiency}\\
       & & (x,y,z) & (s) & & (\%) \\

      \hline
      $256^3$            & 1  & (1,1,1) & 8.93 & -    & -   \\
	  $256^2 \times 512$ & 2  & (1,1,2) & 8.95 & 1.00 & 99.7 \\
      $256 \times 512^2$ & 4  & (1,2,2) & 8.93 & 1.00 & 99.9 \\ 
      $512^3$		    & 8  & (2,2,2) & 8.94 & 1.00 & 99.8 \\
      $1024^3$			& 64 & (4,4,4) & 9.17 & 0.97 & 97.4 \\
      $1024^2 \times2048$& 128 & (4,4,8) & 9.34 & 0.96 & 95.6 \\
      $1024 \times 2048^2$& 256 & (4,8,8) & 9.44 & 0.95 & 94.6 \\
      $2048^3$			  & 512 & (8,8,8) & 9.68 & 0.92 & 92.2 \\
      $2048^3 \times 4096$ & 1024 & (8,8,16) & 9.61 & 0.92 & 92.9 \\
      $4096^3$             & 4096 & (16,16,16) & 9.81 & 0.91 & 91.2\\
    \end{tabular}
    \caption{Weak scaling measurements for fixed box size of $256^3$ per domain are presented above. The wall-clock time corresponds to the time to complete $630$ time-steps (the number of time-steps for a full $256^3$ size simulation). In addition we present a speed-up as well as a parallel efficiency.\label{tab3}}
  \end{center}
\end{table*}

\begin{table*}
  \begin{center}
    \begin{tabular}{c|c|c|c|c|c} 
      \textbf{Box size} & \textbf{Number of GPU's} & \textbf{Domain decomposition} & \textbf{Wall-clock time} & \textbf{Speed-Up} & \textbf{Efficiency}\\
       & & (x,y,z) & (s) & & (\%) \\
             \hline
      $512^3$& 1  & (1,1,1)  & 96.0   & -          & - \\
     $1024^3$& 8   & (2,2,2) & 108.27 & 0.89    & 88.7\% \\
     $2048^3$& 64   & (4,4,4) & 108.97 & 0.88    & 88.1\% \\
     $4096^3$& 512   & (8,8,8) & 117.75 & 0.82    & 81.5\% \\
     $8192^3$& 4096   & (16,16,16) & 124.41 & 0.77    & 77.1\% \\
      \hline
    \end{tabular}
    \caption{Derived weak scaling measurements for fixed box size of $512^3$ per domain are presented above. The wall-clock time corresponds to the time to complete $1270$ time-steps (the number of time-steps for a full $512^3$ size simulation). These are derived from the strong scaling measurements above. In addition we present a speed-up as well as a parallel efficiency.\label{tab4}}
  \end{center}
\end{table*}

However disappointing the strong scaling might be, one can argue that given the relatively short run-times across the board, there isn't a need for good strong scaling \footnote{We remark that this doesn't mean strong scaling is not useful. In fact it is useful to select the most the efficient (ie. less costly in node-hours) number of GPU's for a single simulation}. Indeed, while the largest Abelian-Higgs field theory simulations currently described in the literature have box sizes of $4096^3$, we have been able to carry out full (production run level) $8192^3$ simulations, which using $4096$ GPUs take only $33.2$ minutes of wall clock time. In the same $4096$ GPUs, production runs of size $4096^3$ (the largest reported in the literature so far for cosmic strings) take less than $3$ minutes of wall clock time. 

On the other hand, weak scaling for $256^3$ is almost perfect up to thousands of GPUs with the lowest detected efficiency being of $91\%$---see Table \ref{tab3} and also the middle panels of Figure \ref{fig4}. This strongly suggests the overlap is being successful. Note however, that the overlap is no silver-bullet. As soon as the box size per GPU increases to $512^3$ the overlap isn't enough to completely hide the communications cost, which, hits a lower parallel efficiency (around $77\%$) at the largest process count ($4096$). While the reason for the larger fractional overhead of communications at $512^3$ is so far unknown and warrants further investigation, we can speculate it might be due to overlap combined with automatic data movement that exists between GPU and CPU when using Unified Memory buffers. If it is so, possible solutions include manually moving data and the use of pinned host memory (which can be accessed at will by a GPU). Our code can thus yield production grade simulations with the largest sizes in the literature, $4096^3$, in between 140 and 180 node-hours (the exact number depending on the number of GPUs being used), while for $8192^3$ simulations this increases by a factor of 16 (a factor of 8 due to the increased volume, and 2 due to the increased dynamic range) to about 2200 node-hours. Comparing to a CPU simulation is difficult without having one such simulation, but we can compare with the standard Abelian-Higgs simulation of \cite{Daverio:2015nva,Hindmarsh:2017qff}. We were provided timings via \citet{Hind:email} for a single $4096^3$ run on the older Monte Rosa system for evolution and winding writing. Given that we forego writing windings to disk (choosing only to save the result of the estimator instead), we will compare only with the evolution numbers. A run with $32768$ processes would require, given $32$ cores per node, 1024 nodes to be used for 5.128 hours, i.e. a total of 5 251 node-hours. So we can simulate an 8 times larger lattice for twice as many time-steps ($8192^3$) while spending a bit less than half of the billed node-hours of a traditional $4096^3$ run. In other words, one can say that at this high end of contemporary high-performance computing facilities our code is faster by a factor of 30.

Note that compute time on Piz Daint is book-kept in node-hours, which is why we presented and compared in node-hours in lieu of core-hours.  The other reason we prefer to compare in node-hours instead of core-hours is due to the ambiguity of defining what a “core” might be in a GPU architecture. In order to explain why this is ambiguous let’s follow the definition of a Graphics processing unit as a collection of Streaming Multiprocessors (SM’s). In the case of Pascal GP100 (the architecture of the Tesla cards at Piz Daint), there are $56$ SM’s, each with $64$ single-precision CUDA cores (3584 CUDA cores total). All instructions act upon groups of 32 threads (called warps). One might then assume that only 2 warps can execute at the same time in GP100 SM, but in reality this is not the way to achieve good performance, and therefore it is not what is done in our simulation. The correct way is to oversubscribe and allow multiple warps to be executed at the same time (each doing different instructions). The scheduler of each SM can switch between warps at will, always aiming to hide the cost of latency (meaning hopefully one will have more than 64 threads per SM being executed). How many warps can be in flight will vary from simulation to simulation, depending on the amount of resources (for example, amount of register or shared memory – types of fast memory available to each SM). An analysis for our CUDA kernels can be found in our previous paper \cite{Correia:2018gew}.

The way PRACE converts node-hours to core-hours in Project Access applications is to consider the normal CPU cores (in this case $12$) and the number of SM’s as if they were cores (as explained, not $100\%$ correct), which at Piz Daint gives 68 cores per node. This doesn’t completely punish GPU simulations (considering 3584 CUDA cores would result in a large number of core-hours) but at the same time, it’s not an entirely realistic number. Even using the PRACE definition, there is another source of ambiguity: we do not use all CPU cores of a node, as only one is required to schedule work to the GPU. We could either use the billed core-hours by converting from node-hours to core-hours with a factor of $68$ or define an “effectively used” number of core-hours, using a factor of $57$ ($56$ SM’s + $1$ CPU core).

To end this section we would like to comment on the usage of other metrics to assess application performance in our case. Given that, like most stencil CUDA kernels in the literature with $2.5D$ decomposition, our CUDA kernels end up being memory bound, we do not believe that the amount of floating point operations is a useful metric. It thus follows that we must characterize how memory bound each CUDA kernel is. This was done in the past as seen in \citet{Correia:2018gew} for a different GPU (specifically a QUADRO P5000 at our local cluster facility). Note however that these kernels handled the boundary conditions differently (without including ghost cells as described above). The increase of the sub-domain by two along each direction and the subsequent misaligned access do reduce the memory access bandwidth, but only impact the overall run-time of each kernel slightly (i.e., by $2\%$ in a worst-case-scenario).

\section{\label{concl}Conclusions and outlook}

We have extended our previous GPU-accelerated Abelian-Higgs string evolution code to be able to harness more than on graphical accelerator. To do so we used the Message Passing Interface to handle the necessary boundary terms of a 3D decomposed lattice. Each sub-lattice is evolved on an accelerator with the Compute Unified Device Architecture (CUDA).

In this paper we have validated the code by comparing its outcomes to those described in the literature, and also quantified its scalability. To summarize, the validation confirms a previously noticed slight decrease of the rate of change of the  mean string separation as the box size is increased, while no such effect is seen in the average string velocity. A detailed study of this effect, and its possible relation to the relevant energy loss mechanisms for the cosmic strings, is left for subsequent work.

When it comes to scalability we obtain near-perfect weak scaling up to $4096$ GPUs. Strong scalability is not as good, and from a comparison with the literature \citep{gmd-11-1665-2018} we might expect a processor only version to have a better strong scale behaviour. While this is not a limiting factor for the scientific exploitation of the code, it will be worth to explore techniques to improve this behaviour. One such possibility is the hypercube decomposition of \citet{BlancoPillado:2010sy}, which avoids communication and therefore achieves better scalability. 

There are other possible improvements we can implement on this simulation. One would be the inclusion of Mesh Refinement techniques \citep{Drew:2019mzc,Helfer:2018qgv} which can be used to probe smaller and smaller sub-horizon scales starting from a course lattice and refining as necessary. This would enable probing such small scales without requiring unnecessarily large lattices. Conversely, this also means one can save memory and thus simulate larger comoving sub-domains (albeit with larger spacing) per GPU, enabling larger dynamical ranges. Having a hybrid CPU/GPU simulation in this case would also be an interesting possibility as offloading computations to the GPU would only occur if a sufficiently large amount of threads on a sub-lattice can be used. 
 
 Another important extension would be the exploration of Fast-Fourier Transform capabilities. While one can already use this simulation to obtain observational imprints of string networks by focusing on calibrating semi-analytical modelling (see below for comments on this), it is true that in order to obtain said imprints without modelling one needs to use Fast-Fourier Transforms (FFT's) of the energy-momentum tensor (Unequal Time Correlators). The question of the viability of multiple GPU's for FFT's is however a difficult problem, in particular due to the movement of data often necessary to compute lines of distributed FFT's. Fortunately, some progress has been done on the literature \citep{DBLP:journals/corr/GholamiHMB15} and more recently for the pseudo-spectral solver of \citep{10.1145/3295500.3356209}. In the latter, a careful adaptation of the simulation to the topology of the machine, consideration about on-node and off-node data transfer and compute-communications overlap enable a favourable speed-up to be obtained. We add that the considerations made in that paper to adapt to the topology of Summit, are merely a reflection of the recent trend amongst the top500 \cite{top} for denser nodes (more GPUs and CPU cores per node, less total nodes).

It is also worthy of note that our code is not Input/Output bound when only outputting the network diagnostic quantities every few timesteps, which is the case in standard (production) runs. This changes when outputting an entire lattice of some quantity (such as the absolute value of scalar field, or the windings) every few timesteps for visualization or other detailed diagnostic purposes. As such we are currently exploring in-situ visualization techniques, as previously described in \citet{Ayachit:2015:PCE:2828612.2828624} in order to evade this bottleneck.

In conclusion, this new version of our Abelian-Higgs cosmic string simulation can do production runs for the largest box sizes seen in the literature ($4096^3$) in very competitive amounts of node-hours, and indeed do even larger boxes ($8192^3$) in very reasonable amounts of node-hours. Given the excellent weak scalability at $256^3$ per process, which we have demonstrated up to 4096 GPUs, the numbers are even more appealing when expressed in terms of wall clock time: less than $3$ minutes for $4096^3$, and $33.2$ minutes for $8192^3$ simulations, using the said $4096$ GPUs.
 
We are currently leveraging such an advantage by simulating hundreds of networks at differing expansion rates to calibrate the semi-analytical model of string evolution \citep{Martins:1996jp,Martins:2000cs}, appropriately extending it to account for the correct velocity-dependencies of energy loss and curvature, as seen in \citet{Rybak1}. This was done for small boxes in \citet{Correia:2019bdl}, enabling a first quantitative comparison of the relative importance of the energy loss mechanisms of the networks. Given the changes in the asymptotic quantities seen in Table \ref{tab1}, it will be important to re-calibrate the model for each box size we are currently able to simulate. The availability of a large sample of simulations with increased spatial resolution and dynamic range will also enable a detailed study of the amount of small-scale structure of the network itself. Both of these have obvious implications for any rigorous assessment of the observational consequences of cosmic string networks. 

\section*{Acknowledgements}
This work was financed by FEDER---Fundo Europeu de Desenvolvimento Regional funds through the COMPETE 2020---Operational Programme for Competitiveness and Internationalisation (POCI), and by Portuguese funds through FCT - Funda\c c\~ao para a Ci\^encia e a Tecnologia in the framework of the project POCI-01-0145-FEDER-028987. J.R.C. is supported by an FCT fellowship (grant reference SFRH/BD/130445/2017). We gratefully acknowledge the support of NVIDIA Corporation with the donation of the Quadro P5000 GPU used for this research.

We acknowledge PRACE for awarding us access to Piz Daint at CSCS, Switzerland, through Preparatory Access proposal 2010PA4610 and  Project Access proposal 2019204986. Technical support from Jean Favre at CSCS is gratefully acknowledged.

\bibliography{artigo}

\begin{thebibliography}{43}
\expandafter\ifx\csname natexlab\endcsname\relax\def\natexlab#1{#1}\fi
\providecommand{\url}[1]{\texttt{#1}}
\providecommand{\href}[2]{#2}
\providecommand{\path}[1]{#1}
\providecommand{\DOIprefix}{doi:}
\providecommand{\ArXivprefix}{arXiv:}
\providecommand{\URLprefix}{URL: }
\providecommand{\Pubmedprefix}{pmid:}
\providecommand{\doi}[1]{\href{http://dx.doi.org/#1}{\path{#1}}}
\providecommand{\Pubmed}[1]{\href{pmid:#1}{\path{#1}}}
\providecommand{\bibinfo}[2]{#2}
\ifx\xfnm\relax \def\xfnm[#1]{\unskip,\space#1}\fi
\bibitem[{Abbott et~al.(2018)}]{Abbott:2017mem}
\bibinfo{author}{Abbott, B.P.}, et~al. (\bibinfo{collaboration}{LIGO
  Scientific, Virgo}), \bibinfo{year}{2018}.
\newblock \bibinfo{title}{{Constraints on cosmic strings using data from the
  first Advanced LIGO observing run}}.
\newblock \bibinfo{journal}{Phys. Rev.} \bibinfo{volume}{D97},
  \bibinfo{pages}{102002}.
\newblock \DOIprefix\doi{10.1103/PhysRevD.97.102002},
  \href{http://arxiv.org/abs/1712.01168}{\tt arXiv:1712.01168}.
\bibitem[{Achucarro et~al.(2014)Achucarro, Avgoustidis, Leite, Lopez-Eiguren,
  Martins, Nunes and Urrestilla}]{Semilocals}
\bibinfo{author}{Achucarro, A.}, \bibinfo{author}{Avgoustidis, A.},
  \bibinfo{author}{Leite, A.M.M.}, \bibinfo{author}{Lopez-Eiguren, A.},
  \bibinfo{author}{Martins, C.J.A.P.}, \bibinfo{author}{Nunes, A.S.},
  \bibinfo{author}{Urrestilla, J.}, \bibinfo{year}{2014}.
\newblock \bibinfo{title}{{Evolution of semilocal string networks: Large-scale
  properties}}.
\newblock \bibinfo{journal}{Phys. Rev.} \bibinfo{volume}{D89},
  \bibinfo{pages}{063503}.
\newblock \DOIprefix\doi{10.1103/PhysRevD.89.063503},
  \href{http://arxiv.org/abs/1312.2123}{\tt arXiv:1312.2123}.
\bibitem[{Ade et~al.(2014)}]{Ade:2013xla}
\bibinfo{author}{Ade, P.A.R.}, et~al. (\bibinfo{collaboration}{Planck}),
  \bibinfo{year}{2014}.
\newblock \bibinfo{title}{{Planck 2013 results. XXV. Searches for cosmic
  strings and other topological defects}}.
\newblock \bibinfo{journal}{Astron. Astrophys.} \bibinfo{volume}{571},
  \bibinfo{pages}{A25}.
\newblock \DOIprefix\doi{10.1051/0004-6361/201321621},
  \href{http://arxiv.org/abs/1303.5085}{\tt arXiv:1303.5085}.
\bibitem[{Allen and Shellard(1990)}]{AS}
\bibinfo{author}{Allen, B.}, \bibinfo{author}{Shellard, E.P.S.},
  \bibinfo{year}{1990}.
\newblock \bibinfo{title}{Cosmic string evolution: A numerical simulation}.
\newblock \bibinfo{journal}{Phys. Rev. Lett.} \bibinfo{volume}{64},
  \bibinfo{pages}{119--122}.
\bibitem[{Ayachit et~al.(2015)Ayachit, Bauer, Geveci, O'Leary, Moreland, Fabian
  and Mauldin}]{Ayachit:2015:PCE:2828612.2828624}
\bibinfo{author}{Ayachit, U.}, \bibinfo{author}{Bauer, A.},
  \bibinfo{author}{Geveci, B.}, \bibinfo{author}{O'Leary, P.},
  \bibinfo{author}{Moreland, K.}, \bibinfo{author}{Fabian, N.},
  \bibinfo{author}{Mauldin, J.}, \bibinfo{year}{2015}.
\newblock \bibinfo{title}{Paraview catalyst: Enabling in situ data analysis and
  visualization}, in: \bibinfo{booktitle}{Proceedings of the First Workshop on
  In Situ Infrastructures for Enabling Extreme-Scale Analysis and
  Visualization}, \bibinfo{publisher}{ACM}, \bibinfo{address}{New York, NY,
  USA}. pp. \bibinfo{pages}{25--29}.
\newblock \URLprefix \url{http://doi.acm.org/10.1145/2828612.2828624},
  \DOIprefix\doi{10.1145/2828612.2828624}.
\bibitem[{Bennett and Bouchet(1990)}]{BB}
\bibinfo{author}{Bennett, D.P.}, \bibinfo{author}{Bouchet, F.R.},
  \bibinfo{year}{1990}.
\newblock \bibinfo{title}{High resolution simulations of cosmic string
  evolution. 1. network evolution}.
\newblock \bibinfo{journal}{Phys. Rev.} \bibinfo{volume}{D41},
  \bibinfo{pages}{2408}.
\bibitem[{Bevis et~al.(2007)Bevis, Hindmarsh, Kunz and
  Urrestilla}]{Bevis:2006mj}
\bibinfo{author}{Bevis, N.}, \bibinfo{author}{Hindmarsh, M.},
  \bibinfo{author}{Kunz, M.}, \bibinfo{author}{Urrestilla, J.},
  \bibinfo{year}{2007}.
\newblock \bibinfo{title}{{CMB power spectrum contribution from cosmic strings
  using field-evolution simulations of the Abelian Higgs model}}.
\newblock \bibinfo{journal}{Phys. Rev.} \bibinfo{volume}{D75},
  \bibinfo{pages}{065015}.
\newblock \DOIprefix\doi{10.1103/PhysRevD.75.065015},
  \href{http://arxiv.org/abs/astro-ph/0605018}{\tt arXiv:astro-ph/0605018}.
\bibitem[{Bevis et~al.(2010)Bevis, Hindmarsh, Kunz and
  Urrestilla}]{Bevis:2010gj}
\bibinfo{author}{Bevis, N.}, \bibinfo{author}{Hindmarsh, M.},
  \bibinfo{author}{Kunz, M.}, \bibinfo{author}{Urrestilla, J.},
  \bibinfo{year}{2010}.
\newblock \bibinfo{title}{{CMB power spectra from cosmic strings: predictions
  for the Planck satellite and beyond}}.
\newblock \bibinfo{journal}{Phys. Rev.} \bibinfo{volume}{D82},
  \bibinfo{pages}{065004}.
\newblock \DOIprefix\doi{10.1103/PhysRevD.82.065004},
  \href{http://arxiv.org/abs/1005.2663}{\tt arXiv:1005.2663}.
\bibitem[{Binetruy et~al.(2012)Binetruy, Bohe, Caprini and Dufaux}]{LISA}
\bibinfo{author}{Binetruy, P.}, \bibinfo{author}{Bohe, A.},
  \bibinfo{author}{Caprini, C.}, \bibinfo{author}{Dufaux, J.F.},
  \bibinfo{year}{2012}.
\newblock \bibinfo{title}{{Cosmological Backgrounds of Gravitational Waves and
  eLISA/NGO: Phase Transitions, Cosmic Strings and Other Sources}}.
\newblock \bibinfo{journal}{JCAP} \bibinfo{volume}{1206}, \bibinfo{pages}{027}.
\newblock \DOIprefix\doi{10.1088/1475-7516/2012/06/027},
  \href{http://arxiv.org/abs/1201.0983}{\tt arXiv:1201.0983}.
\bibitem[{Blanco-Pillado et~al.(2011)Blanco-Pillado, Olum and Shlaer}]{Blanco}
\bibinfo{author}{Blanco-Pillado, J.J.}, \bibinfo{author}{Olum, K.D.},
  \bibinfo{author}{Shlaer, B.}, \bibinfo{year}{2011}.
\newblock \bibinfo{title}{{Large parallel cosmic string simulations: New
  results o n loop production}}.
\newblock \bibinfo{journal}{Phys. Rev.} \bibinfo{volume}{D83},
  \bibinfo{pages}{083514}.
\newblock \DOIprefix\doi{10.1103/PhysRevD.83.083514},
  \href{http://arxiv.org/abs/1101.5173}{\tt arXiv:1101.5173}.
\bibitem[{Blanco-Pillado et~al.(2012)Blanco-Pillado, Olum and
  Shlaer}]{BlancoPillado:2010sy}
\bibinfo{author}{Blanco-Pillado, J.J.}, \bibinfo{author}{Olum, K.D.},
  \bibinfo{author}{Shlaer, B.}, \bibinfo{year}{2012}.
\newblock \bibinfo{title}{{A new parallel simulation technique}}.
\newblock \bibinfo{journal}{J. Comput. Phys.} \bibinfo{volume}{231},
  \bibinfo{pages}{98--108}.
\newblock \DOIprefix\doi{10.1016/j.jcp.2011.08.029},
  \href{http://arxiv.org/abs/1011.4046}{\tt arXiv:1011.4046}.
\bibitem[{Briggs et~al.(2014)Briggs, Pennycook, Shellard, Martins, Woodacre and
  Feind}]{Intel}
\bibinfo{author}{Briggs, J.}, \bibinfo{author}{Pennycook, S.J.},
  \bibinfo{author}{Shellard, E.P.S.}, \bibinfo{author}{Martins, C.J.A.P.},
  \bibinfo{author}{Woodacre, M.}, \bibinfo{author}{Feind, K.},
  \bibinfo{year}{2014}.
\newblock \bibinfo{title}{{Unveiling the Early Universe: Optimizing Cosmology
  Workloads for Intel Xeon Phi Coprocessors in an SGI UV20 00 System}}.
\newblock \bibinfo{type}{Technical Report}. SGI/Intel White Paper.
\bibitem[{Correia and Martins(2019)}]{Correia:2019bdl}
\bibinfo{author}{Correia, J.}, \bibinfo{author}{Martins, C.},
  \bibinfo{year}{2019}.
\newblock \bibinfo{title}{{Extending and Calibrating the Velocity dependent
  One-Scale model for Cosmic Strings with One Thousand Field Theory
  Simulations}}.
\newblock \bibinfo{journal}{Phys. Rev. D} \bibinfo{volume}{100},
  \bibinfo{pages}{103517}.
\newblock \DOIprefix\doi{10.1103/PhysRevD.100.103517},
  \href{http://arxiv.org/abs/1911.03163}{\tt arXiv:1911.03163}.
\bibitem[{Correia and Martins(2020a)}]{Correia:2020gkj}
\bibinfo{author}{Correia, J.}, \bibinfo{author}{Martins, C.},
  \bibinfo{year}{2020}a.
\newblock \bibinfo{title}{{Quantifying the effect of cooled initial conditions
  on cosmic string network evolution}}.
\newblock \bibinfo{journal}{Phys. Rev. D} \bibinfo{volume}{102},
  \bibinfo{pages}{043503}.
\newblock \DOIprefix\doi{10.1103/PhysRevD.102.043503},
  \href{http://arxiv.org/abs/2007.12008}{\tt arXiv:2007.12008}.
\bibitem[{Correia and Martins(2017)}]{PhysRevE.96.043310}
\bibinfo{author}{Correia, J.R.C.C.C.}, \bibinfo{author}{Martins, C.J.A.P.},
  \bibinfo{year}{2017}.
\newblock \bibinfo{title}{General purpose graphics-processing-unit
  implementation of cosmological domain wall network evolution}.
\newblock \bibinfo{journal}{Phys. Rev. E} \bibinfo{volume}{96},
  \bibinfo{pages}{043310}.
\newblock \URLprefix \url{https://link.aps.org/doi/10.1103/PhysRevE.96.043310},
  \DOIprefix\doi{10.1103/PhysRevE.96.043310}.
\bibitem[{Correia and Martins(2020b)}]{Correia:2018gew}
\bibinfo{author}{Correia, J.R.C.C.C.}, \bibinfo{author}{Martins, C.J.A.P.},
  \bibinfo{year}{2020}b.
\newblock \bibinfo{title}{{Abelian-Higgs Cosmic String Evolution with CUDA}}.
\newblock \bibinfo{journal}{Astronomy and Computing} \bibinfo{volume}{32},
  \bibinfo{pages}{100388}.
\newblock \DOIprefix\doi{https://doi.org/10.1016/j.ascom.2020.100388},
  \href{http://arxiv.org/abs/1809.00995}{\tt arXiv:1809.00995}.
\bibitem[{Daverio et~al.(2016)Daverio, Hindmarsh, Kunz, Lizarraga and
  Urrestilla}]{Daverio:2015nva}
\bibinfo{author}{Daverio, D.}, \bibinfo{author}{Hindmarsh, M.},
  \bibinfo{author}{Kunz, M.}, \bibinfo{author}{Lizarraga, J.},
  \bibinfo{author}{Urrestilla, J.}, \bibinfo{year}{2016}.
\newblock \bibinfo{title}{{Energy-momentum correlations for Abelian Higgs
  cosmic strings}}.
\newblock \bibinfo{journal}{Phys. Rev. D} \bibinfo{volume}{93},
  \bibinfo{pages}{085014}.
\newblock \DOIprefix\doi{10.1103/PhysRevD.95.049903},
  \href{http://arxiv.org/abs/1510.05006}{\tt arXiv:1510.05006}.
  \bibinfo{note}{[Erratum: Phys.Rev.D 95, 049903 (2017)]}.
\bibitem[{Drew and Shellard(2019)}]{Drew:2019mzc}
\bibinfo{author}{Drew, A.}, \bibinfo{author}{Shellard, E.P.S.},
  \bibinfo{year}{2019}.
\newblock \bibinfo{title}{{Radiation from Global Topological Strings using
  Adaptive Mesh Refinement: Methodology and Massless Modes}}
  \href{http://arxiv.org/abs/1910.01718}{\tt arXiv:1910.01718}.
\bibitem[{Finelli et~al.(2018)}]{CORE}
\bibinfo{author}{Finelli, F.}, et~al. (\bibinfo{collaboration}{CORE}),
  \bibinfo{year}{2018}.
\newblock \bibinfo{title}{{Exploring cosmic origins with CORE: Inflation}}.
\newblock \bibinfo{journal}{JCAP} \bibinfo{volume}{04}, \bibinfo{pages}{016}.
\newblock \DOIprefix\doi{10.1088/1475-7516/2018/04/016},
  \href{http://arxiv.org/abs/1612.08270}{\tt arXiv:1612.08270}.
\bibitem[{Fuhrer et~al.(2018)Fuhrer, Chadha, Hoefler, Kwasniewski, Lapillonne,
  Leutwyler, L\"uthi, Osuna, Sch\"ar, Schulthess and Vogt}]{gmd-11-1665-2018}
\bibinfo{author}{Fuhrer, O.}, \bibinfo{author}{Chadha, T.},
  \bibinfo{author}{Hoefler, T.}, \bibinfo{author}{Kwasniewski, G.},
  \bibinfo{author}{Lapillonne, X.}, \bibinfo{author}{Leutwyler, D.},
  \bibinfo{author}{L\"uthi, D.}, \bibinfo{author}{Osuna, C.},
  \bibinfo{author}{Sch\"ar, C.}, \bibinfo{author}{Schulthess, T.C.},
  \bibinfo{author}{Vogt, H.}, \bibinfo{year}{2018}.
\newblock \bibinfo{title}{Near-global climate simulation at 1\,km resolution:
  establishing a performance baseline on 4888\,gpus with cosmo 5.0}.
\newblock \bibinfo{journal}{Geoscientific Model Development}
  \bibinfo{volume}{11}, \bibinfo{pages}{1665--1681}.
\newblock \URLprefix \url{https://www.geosci-model-dev.net/11/1665/2018/},
  \DOIprefix\doi{10.5194/gmd-11-1665-2018}.
\bibitem[{Gholami et~al.(2015)Gholami, Hill, Malhotra and
  Biros}]{DBLP:journals/corr/GholamiHMB15}
\bibinfo{author}{Gholami, A.}, \bibinfo{author}{Hill, J.},
  \bibinfo{author}{Malhotra, D.}, \bibinfo{author}{Biros, G.},
  \bibinfo{year}{2015}.
\newblock \bibinfo{title}{Accfft: {A} library for distributed-memory {FFT} on
  {CPU} and {GPU} architectures}.
\newblock \bibinfo{journal}{CoRR} \bibinfo{volume}{abs/1506.07933}.
\newblock \URLprefix \url{http://arxiv.org/abs/1506.07933},
  \href{http://arxiv.org/abs/1506.07933}{\tt arXiv:1506.07933}.
\bibitem[{Helfer et~al.(2019)Helfer, Aurrekoetxea and Lim}]{Helfer:2018qgv}
\bibinfo{author}{Helfer, T.}, \bibinfo{author}{Aurrekoetxea, J.C.},
  \bibinfo{author}{Lim, E.A.}, \bibinfo{year}{2019}.
\newblock \bibinfo{title}{{Cosmic String Loop Collapse in Full General
  Relativity}}.
\newblock \bibinfo{journal}{Phys. Rev.} \bibinfo{volume}{D99},
  \bibinfo{pages}{104028}.
\newblock \DOIprefix\doi{10.1103/PhysRevD.99.104028},
  \href{http://arxiv.org/abs/1808.06678}{\tt arXiv:1808.06678}.
\bibitem[{Hindmarsh and Daverio()}]{Hind:email}
\bibinfo{author}{Hindmarsh, M.}, \bibinfo{author}{Daverio, D.}, .
\newblock \bibinfo{howpublished}{Private communication, 20 December 2019}.
\bibitem[{Hindmarsh et~al.(2017a)Hindmarsh, Lizarraga, Urrestilla, Daverio and
  Kunz}]{Hindmarsh:2017qff}
\bibinfo{author}{Hindmarsh, M.}, \bibinfo{author}{Lizarraga, J.},
  \bibinfo{author}{Urrestilla, J.}, \bibinfo{author}{Daverio, D.},
  \bibinfo{author}{Kunz, M.}, \bibinfo{year}{2017}a.
\newblock \bibinfo{title}{{Scaling from gauge and scalar radiation in Abelian
  Higgs string networks}}.
\newblock \bibinfo{journal}{Phys. Rev.} \bibinfo{volume}{D96},
  \bibinfo{pages}{023525}.
\newblock \DOIprefix\doi{10.1103/PhysRevD.96.023525},
  \href{http://arxiv.org/abs/1703.06696}{\tt arXiv:1703.06696}.
\bibitem[{Hindmarsh et~al.(2017b)Hindmarsh, Rummukainen and Weir}]{HindNab}
\bibinfo{author}{Hindmarsh, M.}, \bibinfo{author}{Rummukainen, K.},
  \bibinfo{author}{Weir, D.J.}, \bibinfo{year}{2017}b.
\newblock \bibinfo{title}{{Numerical simulations of necklaces in SU(2)
  gauge-Higgs field theory}}.
\newblock \bibinfo{journal}{Phys. Rev.} \bibinfo{volume}{D95},
  \bibinfo{pages}{063520}.
\newblock \DOIprefix\doi{10.1103/PhysRevD.95.063520},
  \href{http://arxiv.org/abs/1611.08456}{\tt arXiv:1611.08456}.
\bibitem[{Hindmarsh et~al.(2009)Hindmarsh, Stuckey and
  Bevis}]{Hindmarsh:2008dw}
\bibinfo{author}{Hindmarsh, M.}, \bibinfo{author}{Stuckey, S.},
  \bibinfo{author}{Bevis, N.}, \bibinfo{year}{2009}.
\newblock \bibinfo{title}{{Abelian Higgs Cosmic Strings: Small Scale Structure
  and Loops}}.
\newblock \bibinfo{journal}{Phys. Rev. D} \bibinfo{volume}{79},
  \bibinfo{pages}{123504}.
\newblock \DOIprefix\doi{10.1103/PhysRevD.79.123504},
  \href{http://arxiv.org/abs/0812.1929}{\tt arXiv:0812.1929}.
\bibitem[{Kajantie et~al.(1998)Kajantie, Karjalainen, Laine, Peisa and
  Rajantie}]{Kajantie:1998bg}
\bibinfo{author}{Kajantie, K.}, \bibinfo{author}{Karjalainen, M.},
  \bibinfo{author}{Laine, M.}, \bibinfo{author}{Peisa, J.},
  \bibinfo{author}{Rajantie, A.}, \bibinfo{year}{1998}.
\newblock \bibinfo{title}{{Thermodynamics of gauge invariant U(1) vortices from
  lattice Monte Carlo simulations}}.
\newblock \bibinfo{journal}{Phys. Lett.} \bibinfo{volume}{B428},
  \bibinfo{pages}{334--341}.
\newblock \DOIprefix\doi{10.1016/S0370-2693(98)00440-7},
  \href{http://arxiv.org/abs/hep-ph/9803367}{\tt arXiv:hep-ph/9803367}.
\bibitem[{Kibble(1976)}]{Kibble:1976sj}
\bibinfo{author}{Kibble, T.W.B.}, \bibinfo{year}{1976}.
\newblock \bibinfo{title}{{Topology of Cosmic Domains and Strings}}.
\newblock \bibinfo{journal}{J. Phys.} \bibinfo{volume}{A9},
  \bibinfo{pages}{1387--1398}.
\newblock \DOIprefix\doi{10.1088/0305-4470/9/8/029}.
\bibitem[{Lopez-Eiguren et~al.(2017a)Lopez-Eiguren, Lizarraga, Hindmarsh and
  Urrestilla}]{Lopez-Eiguren:2017dmc}
\bibinfo{author}{Lopez-Eiguren, A.}, \bibinfo{author}{Lizarraga, J.},
  \bibinfo{author}{Hindmarsh, M.}, \bibinfo{author}{Urrestilla, J.},
  \bibinfo{year}{2017}a.
\newblock \bibinfo{title}{{Cosmic Microwave Background constraints for global
  strings and global monopoles}}.
\newblock \bibinfo{journal}{JCAP} \bibinfo{volume}{1707}, \bibinfo{pages}{026}.
\newblock \DOIprefix\doi{10.1088/1475-7516/2017/07/026},
  \href{http://arxiv.org/abs/1705.04154}{\tt arXiv:1705.04154}.
\bibitem[{Lopez-Eiguren et~al.(2017b)Lopez-Eiguren, Urrestilla and
  Achucarro}]{Monopoles}
\bibinfo{author}{Lopez-Eiguren, A.}, \bibinfo{author}{Urrestilla, J.},
  \bibinfo{author}{Achucarro, A.}, \bibinfo{year}{2017}b.
\newblock \bibinfo{title}{{Measuring Global Monopole Velocities, one by one}}.
\newblock \bibinfo{journal}{JCAP} \bibinfo{volume}{1701}, \bibinfo{pages}{020}.
\newblock \DOIprefix\doi{10.1088/1475-7516/2017/01/020},
  \href{http://arxiv.org/abs/1611.09628}{\tt arXiv:1611.09628}.
\bibitem[{Martins(2016)}]{Book}
\bibinfo{author}{Martins, C.J.A.P.}, \bibinfo{year}{2016}.
\newblock \bibinfo{title}{Defect Evolution in Cosmology and Condensed Matter:
  Quantitative Analysis with the Velocity-Dependent One-Scale Model}.
\newblock \bibinfo{publisher}{Springer}.
\bibitem[{Martins et~al.(2016)Martins, Rybak, Avgoustidis and
  Shellard}]{Rybak1}
\bibinfo{author}{Martins, C.J.A.P.}, \bibinfo{author}{Rybak, I.Y.},
  \bibinfo{author}{Avgoustidis, A.}, \bibinfo{author}{Shellard, E.P.S.},
  \bibinfo{year}{2016}.
\newblock \bibinfo{title}{{Extending the velocity-dependent one-scale model for
  domain walls}}.
\newblock \bibinfo{journal}{Phys. Rev.} \bibinfo{volume}{D93},
  \bibinfo{pages}{043534}.
\newblock \DOIprefix\doi{10.1103/PhysRevD.93.043534},
  \href{http://arxiv.org/abs/1602.01322}{\tt arXiv:1602.01322}.
\bibitem[{Martins and Shellard(1996)}]{Martins:1996jp}
\bibinfo{author}{Martins, C.J.A.P.}, \bibinfo{author}{Shellard, E.P.S.},
  \bibinfo{year}{1996}.
\newblock \bibinfo{title}{{Quantitative string evolution}}.
\newblock \bibinfo{journal}{Phys. Rev.} \bibinfo{volume}{D54},
  \bibinfo{pages}{2535--2556}.
\newblock \DOIprefix\doi{10.1103/PhysRevD.54.2535},
  \href{http://arxiv.org/abs/hep-ph/9602271}{\tt arXiv:hep-ph/9602271}.
\bibitem[{Martins and Shellard(2002)}]{Martins:2000cs}
\bibinfo{author}{Martins, C.J.A.P.}, \bibinfo{author}{Shellard, E.P.S.},
  \bibinfo{year}{2002}.
\newblock \bibinfo{title}{{Extending the velocity dependent one scale string
  evolution model}}.
\newblock \bibinfo{journal}{Phys. Rev.} \bibinfo{volume}{D65},
  \bibinfo{pages}{043514}.
\newblock \DOIprefix\doi{10.1103/PhysRevD.65.043514},
  \href{http://arxiv.org/abs/hep-ph/0003298}{\tt arXiv:hep-ph/0003298}.
\bibitem[{Martins and Shellard(2006)}]{FRAC}
\bibinfo{author}{Martins, C.J.A.P.}, \bibinfo{author}{Shellard, E.P.S.},
  \bibinfo{year}{2006}.
\newblock \bibinfo{title}{Fractal properties and small-scale structure of
  cosmic string network s}.
\newblock \bibinfo{journal}{Phys. Rev.} \bibinfo{volume}{D73},
  \bibinfo{pages}{043515}.
\newblock \href{http://arxiv.org/abs/astro-ph/0511792}{\tt
  arXiv:astro-ph/0511792}.
\bibitem[{Moore et~al.(2002)Moore, Shellard and Martins}]{Moore:2001px}
\bibinfo{author}{Moore, J.N.}, \bibinfo{author}{Shellard, E.P.S.},
  \bibinfo{author}{Martins, C.J.A.P.}, \bibinfo{year}{2002}.
\newblock \bibinfo{title}{{On the evolution of Abelian-Higgs string networks}}.
\newblock \bibinfo{journal}{Phys. Rev.} \bibinfo{volume}{D65},
  \bibinfo{pages}{023503}.
\newblock \DOIprefix\doi{10.1103/PhysRevD.65.023503},
  \href{http://arxiv.org/abs/hep-ph/0107171}{\tt arXiv:hep-ph/0107171}.
\bibitem[{Olum and Vanchurin(2007)}]{VVO}
\bibinfo{author}{Olum, K.D.}, \bibinfo{author}{Vanchurin, V.},
  \bibinfo{year}{2007}.
\newblock \bibinfo{title}{Cosmic string loops in the expanding universe}.
\newblock \bibinfo{journal}{Phys. Rev.} \bibinfo{volume}{D75},
  \bibinfo{pages}{063521}.
\newblock \href{http://arxiv.org/abs/astro-ph/0610419}{\tt
  arXiv:astro-ph/0610419}.
\bibitem[{Potter et~al.(2017)Potter, Stadel and Teyssier}]{Potter:2016ttn}
\bibinfo{author}{Potter, D.}, \bibinfo{author}{Stadel, J.},
  \bibinfo{author}{Teyssier, R.}, \bibinfo{year}{2017}.
\newblock \bibinfo{title}{{PKDGRAV3: Beyond Trillion Particle Cosmological
  Simulations for the Next Era of Galaxy Surveys}}.
\newblock \bibinfo{journal}{Computational Astrophysics and Cosmology}
  \bibinfo{volume}{4}, \bibinfo{pages}{2}.
\newblock \DOIprefix\doi{10.1186/s40668-017-0021-1},
  \href{http://arxiv.org/abs/1609.08621}{\tt arXiv:1609.08621}.
\bibitem[{PRACE(2017)}]{pracebest}
\bibinfo{author}{PRACE}, \bibinfo{year}{2017}.
\newblock \bibinfo{title}{Best practice guide gpgpu}.
\newblock
  \bibinfo{howpublished}{\url{https://prace-ri.eu/wp-content/uploads/Best-Practice-Guide_GPGPU.pdf}}.
\bibitem[{Press et~al.(1989)Press, Ryden and Spergel}]{PRS}
\bibinfo{author}{Press, W.H.}, \bibinfo{author}{Ryden, B.S.},
  \bibinfo{author}{Spergel, D.N.}, \bibinfo{year}{1989}.
\newblock \bibinfo{title}{{Dynamical Evolution of Domain Walls in an Expanding
  Universe}}.
\newblock \bibinfo{journal}{Astrophys. J.} \bibinfo{volume}{347},
  \bibinfo{pages}{590--604}.
\newblock \DOIprefix\doi{10.1086/168151}.
\bibitem[{Ravikumar et~al.(2019)Ravikumar, Appelhans and
  Yeung}]{10.1145/3295500.3356209}
\bibinfo{author}{Ravikumar, K.}, \bibinfo{author}{Appelhans, D.},
  \bibinfo{author}{Yeung, P.K.}, \bibinfo{year}{2019}.
\newblock \bibinfo{title}{Gpu acceleration of extreme scale pseudo-spectral
  simulations of turbulence using asynchronism}, in:
  \bibinfo{booktitle}{Proceedings of the International Conference for High
  Performance Computing, Networking, Storage and Analysis},
  \bibinfo{publisher}{Association for Computing Machinery},
  \bibinfo{address}{New York, NY, USA}.
\newblock \URLprefix \url{https://doi.org/10.1145/3295500.3356209},
  \DOIprefix\doi{10.1145/3295500.3356209}.
\bibitem[{Wilson(1974)}]{PhysRevD.10.2445}
\bibinfo{author}{Wilson, K.G.}, \bibinfo{year}{1974}.
\newblock \bibinfo{title}{Confinement of quarks}.
\newblock \bibinfo{journal}{Phys. Rev. D} \bibinfo{volume}{10},
  \bibinfo{pages}{2445--2459}.
\newblock \URLprefix \url{https://link.aps.org/doi/10.1103/PhysRevD.10.2445},
  \DOIprefix\doi{10.1103/PhysRevD.10.2445}.
\bibitem[{Xmartlabs(2020)}]{top}
\bibinfo{author}{Xmartlabs}, \bibinfo{year}{2020}.
\newblock \bibinfo{title}{Top 500 list}.
\newblock
  \bibinfo{howpublished}{\url{https://www.top500.org/lists/top500/list/2020/06/}}.

\end{thebibliography}
\end{document}